\documentclass[12pt]{article}
\pdfoutput=1                          
\usepackage{amsmath}
\usepackage{slashed}
\usepackage{amssymb}
\usepackage{epsfig}
\usepackage{graphicx}
\usepackage{multirow}
\usepackage{color}
\usepackage[normal]{subfigure}
\usepackage{rotating}
\usepackage{hyperref}
\usepackage[margin=0.9in]{geometry}
\usepackage[table]{xcolor}
\usepackage{enumitem}
\usepackage[utf8x]{inputenc}
\usepackage[compress,numbers,sort]{natbib}
\usepackage{authblk}
\usepackage{colortbl}
\usepackage{pdflscape}
\definecolor{LightCyan}{rgb}{0.88,1,1}
\definecolor{piggypink}{rgb}{0.99, 0.87, 0.9}

\usepackage{ucs}
\usepackage{amsmath}
\usepackage{amsfonts}
\usepackage{amssymb}
\usepackage{eucal}               
\usepackage{mathrsfs}          
\usepackage[sans]{dsfont}     
\usepackage[Symbol]{upgreek}  

\usepackage{array}             
\usepackage{booktabs}       
\usepackage{colortbl}          
\usepackage{multirow}
\usepackage{longtable}

\usepackage{fancyhdr}

\usepackage[normalem]{ulem}             
\usepackage{cancel}           

\usepackage{slashed}
\usepackage{simplewick}     
\usepackage{feynmf}           

\usepackage{color}
\definecolor{grigio}{cmyk}{0,0,0,0.1}
\definecolor{rosa}{cmyk}{0,0.1,0.1,0.02}
\definecolor{rosino}{cmyk}{0,0.05,0.05,0.02}
\definecolor{rosas}{cmyk}{0,0.3,0.25,0.05}
\definecolor{celeste}{cmyk}{0.1,0,0,0.02}
\definecolor{giallino}{cmyk}{0,0,0.1,0.02}
\definecolor{rosso}{cmyk}{0,1,1,0.4}
\definecolor{rossos}{cmyk}{0,1,1,0.55}
\definecolor{rossoc}{cmyk}{0,1,1,0.2}
\definecolor{blu}{cmyk}{1,1,0,0.3}
\definecolor{blus}{cmyk}{1,1,0,0.5}
\definecolor{bluc}{cmyk}{1,1,0,0.1}
\definecolor{blucc}{cmyk}{0.7,0.5,0,0}
\definecolor{viola}{cmyk}{0,1,0,0.6}
\definecolor{viola2}{cmyk}{0,1,0.2,0.6}
\definecolor{verde}{cmyk}{0.92,0,0.59,0.25}
\definecolor{verdec}{cmyk}{0.92,0,0.59,0.15}
\definecolor{verdes}{cmyk}{0.92,0,0.59,0.4}
\definecolor{verdino}{cmyk}{0.12,0,0.09,0.02}
\definecolor{giallo}{cmyk}{0,0,1,0}
\definecolor{gialloverde}{cmyk}{0.44,0,0.74,0}
\definecolor{Titolo}{rgb}{0.752941176,0.576470588,0.992156863}
\definecolor{altro}{rgb}{0.094117647,0.650980392,0.643137255}
\definecolor{Peanuts}{rgb}{0.2, 0.4, 0.6}
\definecolor{Pean1}{rgb}{0.6, 0.8, 0.4}
\definecolor{BHO}{rgb}{0.2, 0.8, 1}
\definecolor{Daria}{rgb}{0, 0.9412, 0}
\definecolor{UniPi}{rgb}{0.2549, 0.4627, 0.6275}
\definecolor{UniPidue}{rgb}{0.3216, 0.5804, 0.7882}
\definecolor{rossoCP3}{cmyk}{0,.88,.77,.40}
\definecolor{verdeCP3}{rgb}{0.09765625, 0.57421875, 0.1015625}
\definecolor{bluCP3}{rgb}{0, 0.23, 0.67}

\newcommand{\eg}{e.g.~}
\newcommand{\ie}{i.e.~}

\newcommand{\Eq}[1]{Eq.~\eqref{#1}}
\newcommand{\Fig}[1]{Fig.~\ref{#1}}

\newcommand{\Ref}[1]{Ref.~\cite{#1}}           

\newcommand{\Op}{{\cal O}}		
\newcommand{\Lag}{\mathscr{L}}	
\newcommand{\Mel}{\mathscr{M}}	

\newcommand{\ZZ}{\mathbb{Z}}

\newcommand{\Uh}{U(1)_Y}
\newcommand{\SUd}{SU(2)_{\text{L}}}
\newcommand{\SUt}{SU(3)_{\text{c}}}

\newcommand{\cc}{\text{c}}
\newcommand{\hc}{\text{h.c.}}

\newcommand{\LH}{\text{L}}
\newcommand{\RH}{\text{R}}
\newcommand{\TT}{\text{T}}

\newcommand{\tr}{\mathsf{T}}		

\newcommand{\beq}{\begin{equation}}
\newcommand{\eeq}{\end{equation}}
\newcommand{\ud}{\text{d}}

\newcommand{\bol}[1]{\boldsymbol{#1}}

\newcommand{\ER}{E_\text{R}}

\newcommand{\FERMI}{{\sf Fermi}}
\newcommand{\PAMELA}{{\sf PAMELA}}
\newcommand{\HESS}{{\sf H.E.S.S.}}
\newcommand{\MAGIC}{{\sf MAGIC}}
\newcommand{\CTA}{{\sf CTA}}
\newcommand{\PLANCK}{{\sf Planck}}
\newcommand{\LUX}{{\sf LUX}}

\newcommand{\X}{\CMcal{X}}
\newcommand{\DM}{\CMcal{X}^0}
\newcommand{\mDM}{M}

\ifx\pdfoutput\undefined
\usepackage[dvips,bookmarks]{hyperref}    
\else
\usepackage{hyperref}    
\fi
\hypersetup{colorlinks, bookmarksopen, bookmarksnumbered,
citecolor=verdeCP3, linkcolor=bluCP3, pdfstartview=FitH, urlcolor=rossoCP3}

\def\hhref#1{\href{http://arxiv.org/abs/#1}{#1}} 

\begin{document}

\begin{flushright}
{\footnotesize
{\sc DFPD-2015/TH/29}
}
\end{flushright}
\color{black}

\begin{center}
{\Huge\bf Millicharge or Decay: \\ A Critical Take on \\[4mm] Minimal Dark Matter}

\medskip
\bigskip\color{black}\vspace{0.6cm}

{
{\large\bf Eugenio Del Nobile}$\, ^{a, b}$,
{\large\bf Marco Nardecchia}$\, ^c$,
{\large\bf Paolo Panci}$\, ^d$
}
\\[7mm]
{\it $^a$ \href{http://www.pa.ucla.edu/}{Department of Physics and Astronomy, UCLA}, \\ 475 Portola Plaza, Los Angeles, CA 90095, USA}\\[3mm]
{\it $^b$ \href{http://www.dfa.unipd.it/}{Dipartimento di Fisica e Astronomia ``G.~Galilei''}, \\ Universit\`a di Padova and INFN, Sezione di Padova, Via Marzolo 8, 35131 Padova, Italy}\\[3mm]
{\it $^c$ \href{http://www.damtp.cam.ac.uk/}{DAMTP, University of Cambridge}, \\ Wilberforce Road, Cambridge CB3 0WA, United Kingdom}\\[3mm]
{\it $^d$ \href{http://www.iap.fr}{Institut d'Astrophysique de Paris}, UMR 7095 CNRS, Universit\'e Pierre et Marie Curie, \\ 98 bis Boulevard Arago, Paris 75014, France}\\[3mm]
\end{center}

\bigskip

\centerline{\large\bf Abstract}
\begin{quote}
\large
Minimal Dark Matter (MDM) is a theoretical framework highly appreciated for its minimality and yet its predictivity. Of the two only viable candidates singled out in the original analysis, the scalar eptaplet has been found to decay too quickly to be around today, while the fermionic quintuplet is now being probed by indirect Dark Matter (DM) searches. It is therefore timely to critically review the MDM paradigm, possibly pointing out generalizations of this framework. We propose and explore two distinct directions. One is to abandon the assumption of DM electric neutrality in favor of absolutely stable, millicharged DM candidates which are part of $\SUd$ multiplets with integer isospin. Another possibility is to lower the cutoff of the model, which was originally fixed at the Planck scale, to allow for DM decays. We find new viable MDM candidates and study their phenomenology in detail.
\end{quote}

\newpage

\section{Introduction}
The presence of Dark Matter (DM) in the universe is a clear evidence for new physics beyond the Standard Model (SM). Despite lacking a unique description of DM in terms of elementary particles, a number of general requirements have been identified for DM candidates to fit observations. One of these is stability on cosmological scales.

Stability may be explained in terms of symmetries. One may impose a symmetry on a DM model {\em by hand} to force stability of the DM candidate, hoping this symmetry can be later justified naturally in ultraviolet completions of the model. Another elegant way to ensure stability is instead through accidental symmetries, the same mechanism that makes the proton stable in the SM. In fact, if one considers only local symmetries as fundamental, other exact or approximate global symmetries can arise as accidental ``gifts'' given by the specific matter content of the model, which are preserved up to a certain dimension in an effective theory description.

This is the main idea behind the ``Minimal Dark Matter'' (MDM) setup first presented in Ref.~\cite{Cirelli:2005uq}, where the SM is augmented with a new generic multiplet $\X$ with mass $\mDM$ and quantum numbers $(\bol{c}, \bol{n}, Y)$ under the SM gauge group $\SUt \times \SUd \times \Uh$, without introducing new symmetries. The requirement that the multiplet contains a suitable DM candidate with the correct relic abundance, which is stable on cosmological time scales and is not excluded by present observations, is then used to constrain $\X$'s quantum numbers. For example, the MDM multiplet must be color neutral to avoid the stringent constraints on colored particles~\cite{Starkman:1990nj, Taoso:2007qk} which seem to exclude the parameter space of thermal relics. Moreover, $\bol{n}$ must be odd or $\X$'s components would all have sizable tree-level interactions with the $Z$ boson, which are excluded by direct DM searches. The authors of Ref.~\cite{Cirelli:2005uq} then go on and {\em assume} DM electric neutrality, which implies $Y = 0$. In order to avoid Yukawa couplings with SM fields, as well as dimension-$5$ effective operators that would cause the DM to decay quickly, it must be $\bol{n} \geqslant 5$ for spin-$\frac{1}{2}$ multiplets and $\bol{n} \geqslant 7$ for scalars. Finally, the consistency condition that the theory does not produce a Landau pole below the assumed cut-off at the Planck scale is used to set an upper bound on $\bol{n}$, $\bol{n} \leqslant 5$ for Majorana fermions and $\bol{n} \leqslant 7$ for real scalars\footnote{\label{LandauPoles} We refer the reader to Table 9 of Ref.~\cite{DiLuzio:2015oha} for a list of two-loop Landau poles in theories where the SM is augmented with an extra multiplet with zero hypercharge. Notice that the results presented there are determined by ``integrating in'' the extra multiplet at $\mDM$ equal to the $Z$-boson mass. The case of arbitrary multiplet mass can be derived by considering that the Landau pole scales approximately linearly with $\mDM$. For scalar multiplets, an exhaustive analysis should include also the renormalization of quartic scalar interactions~\cite{Hamada:2015bra}.} (these bounds are conservative with respect to those for Dirac fermions and complex scalars). As a result, the authors of Ref.~\cite{Cirelli:2005uq} single out a fermionic $\SUd$ quintuplet and a scalar septuplet as the only viable MDM multiplets.

The eptaplet candidate was recently excluded by the presence of an overlooked dimension-$5$ operator trilinear in $\X$~\cite{DiLuzio:2015oha} which makes the DM candidate decay too quickly, while the fermionic quintuplet is seriously constrained by gamma-ray line searches in the Galactic Center~\cite{Cirelli:2015bda, Garcia-Cely:2015dda}. In the light of these recent results and of the good sensitivity of present searches to MDM candidates, a critical reanalysis of the MDM framework is timely. Despite the extended literature on the subject and variations thereof (see \eg Refs.~\cite{Boucenna:2015haa, Cheung:2015mea, Heeck:2015qra, Garcia-Cely:2015quu, Culjak:2015qja, Ostdiek:2015aga, Fabbrichesi:2015bta, Antipin:2015xia, Buen-Abad:2015ova, Cai:2015kpa, Cai:2015joa, Hamada:2015fma, Ahriche:2015wha, Aoki:2015nza} for some very recent works), some assumptions and basic aspects of the MDM setup remain, that could be more thoroughly examined, and others that could be easily generalized or naturally extended. These are, for instance, the assumption that the cutoff of the theory is at the Planck scale; or the choice of taking small $\X$-Higgs quartic couplings; or even the seemingly natural assumption that the DM is electrically neutral. The aim of this work is to examine these aspects in detail, proposing generalizations and studying their phenomenological consequences, in the spirit of Gell-Mann's Totalitarian Principle ``everything not forbidden is compulsory''. In extending the MDM framework we find a new DM candidate, a Dirac $\SUd$ triplet, with a larger degree of compatibility with present bounds with respect to the standard MDM quintuplet. However, the constraints are not necessarily relaxed for the other candidates we propose, meaning that most of the scenarios we study explicitly can be ideally probed by present or near-future experiments.

After critically reviewing the MDM setup and its assumptions in Section~\ref{Critical review}, we explore two main directions in Section~\ref{sec:Millicharge} and \ref{Decaying MDM}. In Section~\ref{sec:Millicharge} we abandon the assumption of DM electric neutrality, thus allowing multiplets to get non-zero (although phenomenologically small) hypercharge. These candidates feature a millicharged DM particle which is absolutely stable due to electric charge conservation. This removes the need of a very high cutoff and opens the possibility of large $\SUd$ representations without having to worry about Landau poles. We discuss the phenomenology of these candidates and compute the mass needed to achieve thermal production for few of them. Interestingly, the millicharged $\SUd$ fermionic triplet is found not (yet) to suffer from the stringent gamma-ray line constraints afflicting the standard fermionic quintuplet. In Section~\ref{Decaying MDM} we explore the consequences of lowering the cutoff from the Planck scale, so that the MDM fermionic quintuplet decays with observable consequences in the gamma-ray sky. We compute in detail the photon flux (both continuum and line-like features) from DM decays and constrain the cutoff $\Lambda$ using \FERMI~data on the diffuse isotropic flux and \HESS~data on gamma-ray lines. Were a clear photon line from this candidate's annihilations to be soon detected, gamma-ray data could also be used to gain insight on the scale of new physics $\Lambda$ above the DM mass. We conclude in Section~\ref{Conclusions}. One interesting technical aspect of our work concerns the presence (or absence) in the Lagrangian of operators of the form $\X^3$ or $\X^3 H^2$, which cause the DM to decay quickly. We study the issue with the method of Hilbert series in Appendix~\ref{app:trilinear}, while in Appendix~\ref{app:rates} we give analytic expressions for the total and differential decay rates of the fermionic MDM quintuplet at dimension $6$ in the effective Lagrangian. Finally, we give an analytic treatment of the phase space for $4$-body decays into massless particles for our case of interest in Appendix~\ref{app:phasespace}.

\section{Minimal Dark Matter, a critical review}
\label{Critical review}
As explained in the Introduction, the MDM setup features the addition of an extra multiplet $\X$ to the SM, with quantum numbers $(\bol{c}, \bol{n}, Y)$ under the SM gauge group $\SUt \times \SUd \times \Uh$. Further requirements characterizing the MDM framework~\cite{Cirelli:2005uq}, reported and individually commented below, allow to reduce the number of suitable candidates to a few. For the sake of restricting our discussion to the phenomenologically viable candidates, let us anticipate here some consequences of the requirement that the DM candidate still be allowed by DM searches, see point \ref{cond4} below.

Given the stringent constraints on colored particles~\cite{Starkman:1990nj, Taoso:2007qk} which seem to exclude the parameter space of thermal relic DM~\cite{Cirelli:2005uq}, we restrict ourselves to color-neutral multiplets, $\bol{c} = \bol{1}$. $\bol{n}$ must be odd for $\X$ to contain a viable DM candidate, with no sizable tree-level interactions with the $Z$ boson and the photon which are excluded by direct DM searches. We do not enforce electric neutrality for the DM at this point, so that $Y$ is allowed to take non-zero (but nevertheless very small, see Section~\ref{sec:Millicharge}) values. Fermion multiplets are taken to be vector-like so that they can be given an invariant mass term in the Lagrangian and to cancel anomalies. Notice that the $(\bol{1}, \bol{n}, Y)$ representation with odd $\bol{n}$ is real for $Y = 0$, while it is complex for $Y \neq 0$.

Barring Yukawa interactions of $\X$ with two SM fields, which are explicitly forbidden (by gauge invariance) in the MDM setup to avoid DM decay, the renormalizable Lagrangian of the model is
\begin{align}
\Lag_{Y = 0} &= \Lag_\text{SM} +
\begin{cases}
\frac{1}{2} \overline{\X} i \slashed{D} \X + \frac{\mDM}{2} \X^\tr C^{-1} \X
& \text{for Majorana $\X$},
\\
\frac{1}{2} (D^\mu \X)^\dagger (D_\mu \X) - \frac{\mDM^2}{2} \X^\tr \X - V(\X, H) & \text{for real scalar $\X$},
\end{cases}
\\
\intertext{for $Y = 0$, and}
\Lag_{Y \neq 0} &= \Lag_\text{SM} +
\begin{cases}
\overline{\X} (i \slashed{D} - \mDM) \X & \text{for Dirac $\X$},
\\
(D^\mu \X)^\dagger (D_\mu \X) - \mDM^2 \X^\dagger \X - V(\X, H) & \text{for complex scalar $\X$},
\end{cases}
\end{align}
for $Y \neq 0$, where $C$ is the charge conjugation matrix and $V(\X, H)$ denotes $\X$'s potential plus possible $\X$-$H$ interaction operators. The lightest state contained in the $\X$ multiplet (the DM candidate) is stable under a $\ZZ_2$ symmetry transforming $\X \to - \X$ for $\X$ in a real representation ($Y = 0$), or a $U(1)$ symmetry transforming $\X \to e^{i \theta} \X$ for $\X$ in a complex representation ($Y \neq 0$). $\bol{n}$ and $Y$ are chosen so that no renormalizable and dimension-$5$ interactions exist, that spoil this symmetry thus inducing fast DM decays. This dictates the absence of Yukawa interactions and restricts the operators entering $V(\X, H)$.

The MDM setup is characterized by the following requirements, that allow to further reduce the number of suitable candidates~\cite{Cirelli:2005uq}. After stating each requirement (written in boldface below), we critically review its implications and comment upon possible generalizations.
\begin{enumerate}
\item {\bf ``The  lightest  component  is  automatically  stable  on  cosmological  time-scales''.} \!\!\!\!\!\!
\label{cond1}
\\
The easiest way to satisfy this condition is probably assuming a very light $\X$, so that the DM particle cannot decay to anything. However, such a multiplet would have been already discovered if charged under strong or weak interactions. The only possibility of having a light DM particle seems therefore for $\X$ to have quantum numbers $(\bol{1}, \bol{1}, 0)$ or $(\bol{1}, \bol{1}, \epsilon)$ with very small (but positive\footnote{Here and in the following we take $\epsilon$ to be a positive number. The case of negative hypercharge is trivially related to that of positive hypercharge.}) $\epsilon$. The former, for a real scalar $\X$, is the well studied scalar singlet DM, deemed to be one of the simplest DM models (see \eg Refs.~\cite{Cline:2013gha, Beniwal:2015sdl} and references therein). UV-complete models of fermionic singlet DM usually also feature a scalar messenger connecting the dark and visible sectors, see \eg Ref.~\cite{Kim:2008pp}, since there exist no renormalizable interactions of a fermionic singlet alone with SM fields (the lowest-order interaction of this DM candidate is through the Higgs-portal dimension-$5$ operator $\overline{\X} \X H^\dagger H$, see \eg Refs.~\cite{Fedderke:2014wda, Beniwal:2015sdl} for recent references). The $(\bol{1}, \bol{1}, \epsilon)$ candidate is the so-called millicharged DM. We defer a further examination of this candidate to Section~\ref{sec:Millicharge}.

For a heavy multiplet, the stability condition implies that the only acceptable quantum numbers are those for which an accidental symmetry exists, protecting the DM from decay. Assuming the cutoff $\Lambda$ of the theory to be the Planck scale, as in the original MDM setup, this symmetry must be respected both by renormalizable interactions and by dimension-$5$ effective operators, in order to avoid fast decay of the DM candidate. Decays induced by higher-dimensional operators violating the accidental symmetry have a negligible impact on the DM phenomenology, but can become important if a lower cutoff is assumed.

While it is natural to assume a Planck-scale cutoff, new physics is required to explain experimentally observed phenomena like neutrino oscillations or the matter-antimatter asymmetry in the universe, that may occur at a much lower scale. With no further assumption added to the MDM paradigm, one cannot prevent the new physics responsible for these phenomena to break the accidental symmetry stabilizing the DM. It is then natural to study the effect of higher-order symmetry breaking operators and the phenomenology of decaying MDM as a function of the cutoff scale $\Lambda$. We will discuss all this for the MDM fermionic quintuplet in Sec.~\ref{Decaying MDM}. The scalar eptaplet decays too quickly, as discussed in the following, and it is therefore not a good MDM candidate.

When considering DM decays, the most obvious operators to consider are those that are linear in the DM field, which break the accidental symmetry with the minimum number of $\X$ fields. However, one should worry about all symmetry-breaking operators, including those with a larger number of $\X$ fields.
In particular, we show in Appendix \ref{app:trilinear} that for any $\SUd$ $\bol{n}$-plet $\X$ with integer weak isospin $I = (\bol{n} - 1) / 2$, three $\X$ can be uniquely combined into an $\SUd$ singlet for even $I$ (\ie $\bol{n} = 1, 5, 7, \dots$) or into a triplet for odd $I$ (\ie $\bol{n} = 3, 9, \dots$). Therefore all scalar $(\bol{1}, \bol{n}, 0)$ multiplets with odd $\bol{n}$ allow for dimension-$5$ symmetry-breaking operators of the form $\X^3 H^2$, with $H^2$ either a $\SUd$ singlet or triplet (notice that the $\ZZ_2$ symmetry protecting scalar DM from decay is already broken at dimension $3$ by the operator $\X^3$ for multiplets with even $I$). Upon closing two of the three $\X$ legs in a loop (see \eg Fig.~9 of Ref.~\cite{DiLuzio:2015oha}), these operators induce fast DM decays even assuming a Planck-scale cutoff.

The argument just presented is very general and can be also applied outside the MDM framework. In fact, it concerns any model featuring a color and hypercharge-neutral scalar multiplet containing a DM candidate, unless ad-hoc symmetries are introduced to prevent DM from decaying. A possibility to bypass this drawback of scalar DM could be to assign the scalar multiplet a tiny hypercharge. This generalization of the MDM paradigm will be the subject of the next section.

A similar argument as above can be applied to fermion $(\bol{1}, \bol{n}, 0)$ multiplets with odd $\bol{n}$, for which there exists always a dimension-$7$ symmetry-breaking operator of the type $\X^3 L H$. Cosmological bounds on the DM lifetime, $\tau_\text{DM} > 150 \text{ Gyr} \approx 5 \times 10^{18}$ s~\cite{Audren:2014bca, Aubourg:2014yra}, then fix a minimum cutoff scale that can be estimated using naive dimensional analysis:
\beq
\frac{1}{\tau_\text{DM}} \simeq \frac{1}{16 \pi^2} \frac{\mDM^7}{\Lambda^6} \ ,
\eeq
implying $\Lambda \gtrsim 10^{11}$ GeV for $M \approx 10$ TeV. This bound in turn allows to fix an upper limit on $\bol{n}$ with the requirement that the model has no Landau poles below this minimal cutoff, $\bol{n} \leqslant 5$ (see footnote \ref{LandauPoles}). As for the scalars, the existence of symmetry-breaking operators relies on the multiplet having zero hypercharge as assumed in the original MDM setup; in the next section we will relax this assumption and show that MDM with $Y \neq 0$ can be phenomenologically viable.
\item {\bf ``The  only  renormalizable  interactions  of  $\X$  to  other  SM  particles  are  of  gauge  type,  such  that  new  physics  is  determined  by  one  new  parameter:  the  tree-level  mass  $M$  of  the  Minimal  Dark  Matter  (MDM)  multiplet''.}
\label{cond2}
\\
In line with Gell-Mann's Totalitarian Principle, this condition cannot be satisfied for Lorentz scalars (as already noticed in Ref.~\cite{Cirelli:2005uq}). In fact, the dimension-$4$ operators $\X^\dagger \X H^\dagger H$ and $\X^\dagger t^a_\X \X \, H^\dagger t^a_H H$ with $t^a$ the $\SUd$ gauge group generators in the proper representation, cannot be forbidden by any choice of $\X$'s quantum numbers. If taken into account, these couplings can affect the annihilation cross section, which is relevant for the computation of the relic abundance and thus for determining $\mDM$. The operator $\X^\dagger t^a_\X \X \, H^\dagger t^a_H H$ also affects the mass splitting between the different components of the multiplet. The requirement that the splitting is determined only by loop corrections (see next point) constrains its coefficient to be smaller than $\Op(M / 100 \text{ TeV})$~\cite{Cirelli:2005uq}. Moreover, it was found in Ref.~\cite{Hamada:2015bra} that the renormalization group evolution of quartic couplings in $V(\X, H)$ generates a Landau pole below the Planck scale, and below the Landau pole due to running of the gauge couplings~\cite{DiLuzio:2015oha}, even if the coupling constants are set to zero at the scale $M$. Although the presence of these operators spoils the minimality of the model by introducing extra free parameters, their effect must be included in any truly generic analysis of scalar MDM candidates.
\item {\bf ``Quantum  corrections  generate  a  mass  splitting  $\Delta  \mDM$  such  that  the  lightest  component  of  $\X$  is  neutral.  We  compute  the  value  of  $\mDM$  for  which  the  thermal  relic  abundance  equals  the  measured  DM  abundance''.}
\label{cond3}
\\
If condition \ref{cond2} is met, then the splitting can only be radiative, and its size is fixed by $\X$'s quantum numbers. In this case, as shown in Ref.~\cite{Cirelli:2005uq}, the lightest component of $\X$ is electrically neutral as long as $Y = 0$. Even letting the hypercharge take non-zero values, although small enough to be allowed by DM searches as required by condition \ref{cond4} below, the lightest state is a viable DM candidate albeit electrically charged. Therefore, a lightest state with small $|Y|$ is automatically obtained when conditions \ref{cond2} and \ref{cond4} are enforced simultaneously.
\item {\bf ``The  DM  candidate  is  still  allowed  by  DM  searches''.}
\label{cond4}
\\
As anticipated above, thermal-relic colored DM seems to be excluded by the stringent constraints on strongly-interacting DM~\cite{Starkman:1990nj, Taoso:2007qk, Cirelli:2005uq}. Moreover, constraints from direct DM searches imply that interactions with the photon and the $Z$ boson must be suppressed. This only leaves open the possibility of $(\bol{1}, \bol{n}, Y)$ MDM with odd $\bol{n}$ and either $Y = 0$ or $Y = \epsilon$ with very small but positive $\epsilon$.
\end{enumerate}

Summarizing, we confirmed that, of the two MDM candidates which were so far considered to be viable, the scalar eptaplet is actually ruled out~\cite{DiLuzio:2015oha}. This singles out the fermionic quintuplet as the only viable MDM candidate. We also proposed to extend the MDM framework by separately dropping two of the original assumptions, namely the assumption of a Planck-scale cutoff and the assumption of DM electric charge neutrality. The assumption of a Planck-scale cutoff can be relaxed in favor of a generic cutoff, which then enters the model as a new free parameter that can be probed by studying cosmic-ray signatures of DM decays. We will do that in Section~\ref{Decaying MDM}. Finally, by lifting the hypothesis of electric neutrality of the DM we established the existence a new class of MDM models featuring millicharged DM. We explore this possibility in Section~\ref{sec:Millicharge}.

\section{Millicharged MDM: $(\bol{1}, \bol{n}, \epsilon)$ candidates}
\label{sec:Millicharge}
In this section we explore the possibility of MDM candidates with small hypercharge, $(\bol{1}, \bol{n}, \epsilon)$ with $\epsilon \neq 0$. As in the standard MDM scenario, $\bol{n}$ must be odd to avoid tree-level interactions of the DM particle with the $Z$ boson. Notice that DM-Higgs interactions can induce a mass splitting in the DM components that makes the DM-nucleus scattering at direct detection experiments inelastic, thus drastically reducing the scattering rate. In this case the stringent bounds from direct DM searches become ineffective and relatively large hypercharge assignments are possible, see Refs.~\cite{Hisano:2014kua, Nagata:2014aoa}. We do not pursue this direction here, and stick to small $\epsilon$.

An important feature of these candidates is that the DM has electric charge equal to $\epsilon$ (in units of $e$), and this makes it absolutely stable. In fact, its stability is protected to all orders in the effective field theory expansion by electric charge conservation. What is usually an unwanted feature in a DM candidate, \ie electric charge, is here what stabilizes the multiplet making it a potentially successful candidate!

Since the DM is stable to all orders, one does not need to worry any more about cutoffs. In the original MDM setup, large multiplets were discarded because the presence of Landau poles in the running of the electroweak gauge couplings could indicate new physics that may spoil the accidental symmetry stabilizing the DM. Millicharged DM being absolutely stable now allows to consider, in principle, even large $\bol{n}$'s. In this case, a criterion for setting an upper bound on $\bol{n}$ could be computability. For example we may require that the $1$-loop amplitude does not exceed the tree-level result: roughly speaking, $(\alpha_2 / 4 \pi) G < 1$ with $\alpha_2$ the $\SUd$ fine structure constant and $G$ a $\bol{n}$-dependent group factor. 

Although it may seem odd to consider a field with such a small hypercharge, there is no a priori reason to exclude this possibility: in fact, this choice is allowed by gauge symmetry, and gauge anomaly cancellation is unaffected as long as fermion DM candidates are vector-like.

From a GUT standpoint, one may object that small values of hypercharge are difficult to accommodate in models of grand unification. While this is definitely true, we note that the whole MDM framework is not particularly GUT friendly, since its large multiplets badly modify the running of the SM gauge couplings and moreover they supposedly require a large GUT representation to embed the $\X$ field,\footnote{For example, in $SU(5)$ GUT, the lowest-dimensional irreducible representation containing $(\bol{1}, \bol{5}, 0)$ is $\bol{200}$.} thus generating a severe doublet-triplet splitting-like problem.

There is also a more theoretical advantage that is worth commenting. According to the no-hair theorem~\cite{Bekenstein:1971hc, Bekenstein:1972ky}, gravitational effects break global but not local symmetries. As we observed above, stability of millicharged DM is guaranteed by a local symmetry, the unbroken $U(1)_\text{EM}$. Remarkably, this is the {\em only} symmetry that could be used to {\em completely} stabilize the DM without extending the gauge group of the SM model. This being said, for phenomenological purposes it is enough for a global symmetry stabilizing the DM to be accidentally preserved up to dimension $5$ in an effective theory expansion: in fact, the effects of breaking that symmetry at the Planck scale are small enough to guarantee the stability of the DM on cosmological timescales.

In the following we first review the most stringent constraints on the DM electric charge $\epsilon$, and then discuss the possible millicharged MDM candidates $(\bol{1}, \bol{n}, \epsilon)$ and compute the mass of few of them.

\subsection{Constraints}
Constraints on heavy millicharged particles are inferred from cosmological and astrophysical observations as well as direct laboratory tests~\cite{McDermott:2010pa, Dolgov:2013una, Kouvaris:2013gya}. The most stringent upper bounds on $\epsilon$, summarized in the following, are shown in the right panel of Fig.~\ref{fig:Masses} below. A conceivable lower bound could be obtained by considering the weak gravity conjecture~\cite{ArkaniHamed:2006dz}, which requires $\epsilon > \mDM / M_\text{Pl}$.

\subsubsection{Bounds from CMB}
Millicharged DM particles scatter off electrons and protons at the recombination epoch via Rutherford-like interactions. It was shown that if millicharged particles couple tightly to the baryon-photon plasma during the recombination epoch, they behave like baryons thus affecting the CMB power spectrum in several ways. This kind of bounds were derived by different groups~\cite{McDermott:2010pa, Dolgov:2013una}. In particular, Ref.~\cite{Dolgov:2013una} found that in order to avoid the tight-coupling condition the DM millicharge must be
\beq\label{tight coupling}
\epsilon \lesssim 2.24 \times 10^{-4} \left( \frac{M}{1 \text{ TeV}} \right)^{1/2}
\eeq
for a DM particle much heavier than the proton.

\subsubsection{Direct searches}
Millicharged DM scatters off nuclei via Rutherford-like interactions. In the non-relativistic limit the differential cross section for DM scattering off a nuclear target $T$ with mass $m_T$ and electric charge $e Z_T$ is given by~\cite{Fornengo:2011sz, Panci:2014gga}
\beq
\frac{\ud \sigma_T}{\ud \ER}(v, q^2) = 8 \pi m_T \frac{\alpha^2 \epsilon^2}{v^2 \, q^4} Z_T^2 \, F_T^2(q^2) \ .
\eeq
Here $\ER$ is the nuclear recoil energy, related to the momentum transfer $q$ by $q^2 = 2 m_T \ER$, and $\alpha$ is the electromagnetic fine structure constant. $F_T(q^2)$ is the nuclear Helm form factor~\cite{Helm:1956zz, Lewin:1995rx}, which takes into account the loss of coherence of the interaction at large $q$. Since the interaction is spin-independent, the most stringent bound to date is set by the \LUX~experiment~\cite{Akerib:2013tjd}. We use the tools in Ref.~\cite{DelNobile:2013sia} to infer a $90\%$ CL bound on $\epsilon$ from \LUX. For $M \gtrsim 100$ GeV, only values
\beq
\epsilon \lesssim 7.6 \times 10^{-10} \left( \frac{M}{1 \text{ TeV}} \right)^{1/2}
\eeq
are allowed by \LUX~with $90\%$ confidence. Notice this bound does not apply in the range
\beq
9 \times 10^{-9} \left( \frac{M}{1 \text{ TeV}} \right) \lesssim \epsilon \lesssim 1.1 \times 10^{-2} \left( \frac{M}{1 \text{ TeV}} \right)^{1/2} \ ,
\eeq
because for these values millicharged particles have been evacuated from the galactic disk by supernova explosion shock waves, and galactic magnetic fields prevent them from entering back~\cite{Chuzhoy:2008zy, McDermott:2010pa}. For $\epsilon$ respecting \Eq{tight coupling}, we do not expect DM self-scattering to sufficiently randomize the direction of motion of DM particles before they gyrate out of the disk~\cite{Foot:2010yz}.

These constraints, depicted in the right panel of Fig.~\ref{fig:Masses}, allow for relatively large values of $\epsilon$, which may give rise to interesting phenomenology of millicharged DM candidates. However, for values below the \LUX~bound this parameter does not contribute to the DM phenomenology and can be safely ignored, the only relevant effect being the doubling of the number of $\X$'s degrees of freedom due to passing from a real to a complex representation of the gauge group.

In the following we discuss the possible millicharged candidates and their phenomenology. We first consider $(\bol{1}, \bol{1}, \epsilon)$ candidates, which do not have weak interactions, and then we focus on $(\bol{1}, \bol{n}, \epsilon)$ with $\bol{n} \geqslant 3$.

\subsection{$(\bol{1}, \bol{1}, \epsilon)$ Dirac fermion}
This candidate has only electromagnetic interactions at the renormalizable level. Ref.~\cite{McDermott:2010pa} showed that the parameter space where the DM can be produced thermally with the correct relic abundance is ruled out by observations, most notably by the CMB bound commented above. Therefore, without introducing non-renormalizable interaction, this candidate must be non-thermally produced. Since the details of this production mechanism are highly model dependent, we do not explore this possibility further. Another possibility is to assume a low enough cutoff so that production of this candidate can occur through the dimension-$5$ Higgs-portal operator $\overline{\X} \X H^\dagger H$. In this case, the assumption of thermal production fixes the cutoff as a function of the DM mass, which remains as a free parameter. This candidate has been widely studied in the literature (see \eg Refs.~\cite{Fedderke:2014wda, Beniwal:2015sdl} for recent references), therefore we do not dwell further on this possibility.

\subsection{$(\bol{1}, \bol{1}, \epsilon)$ complex scalar}
With no interactions other than electromagnetic, thermal production of this candidate is ruled out on the same ground as the fermionic $(\bol{1}, \bol{1}, \epsilon)$ candidate. However, as already mentioned, scalar DM can interact with the Higgs through the Higgs portal $\X^\dagger \X H^\dagger H$, which opens a new window for thermal production. Given the strong bounds on $\epsilon$, this candidate behaves basically as a complex scalar field which is completely neutral under the SM. In the assumption we can neglect the quartic self-coupling $(\X^\dagger \X)^2$, the real and complex components of $\X$ do not interact with each other (see below) and can be therefore treated independently as two degenerate real scalar DM particles. Real scalar DM is considered one of the simplest models of DM, and has been widely studied in the literature (see \eg Refs.~\cite{Cline:2013gha, Beniwal:2015sdl} and references therein). The most stringent constraints on thermal relic DM come from the Higgs's invisible decay width~\cite{Belanger:2013xza}, which excludes DM masses below $\sim 50$ GeV, and the \LUX~bound~\cite{Akerib:2013tjd}, which excludes masses from about $10$ GeV to roughly $200$ GeV~\cite{Han:2015hda}, except for a very narrow (few GeV wide) interval around half the Higgs boson mass ($M \approx 60$ GeV) where the annihilation cross section is resonantly enhanced. As shown in Ref.~\cite{Cline:2013gha}, DM masses above $200$ GeV will be probed in the near future by both direct and indirect detection experiments, most notably {\sf XENON1T}~\cite{Aprile:2012zx, Aprile:2015uzo} and \CTA~\cite{Consortium:2010bc}.

\subsection{$(\bol{1}, \bol{n}, \epsilon)$ with $\bol{n} \geqslant 3$}
For $\epsilon$ satisfying the bounds on millicharged DM presented above, DM particles interact mainly with massive gauge bosons. Therefore, the phenomenology of a $(\bol{1}, \bol{n}, \epsilon)$ multiplet is basically identical to that of $(\bol{1}, \bol{n}, 0)$. The only difference, for odd $\bol{n}$, is that $(\bol{1}, \bol{n}, 0)$ is a real representation of the gauge group while $(\bol{1}, \bol{n}, \epsilon)$ is complex. This implies a doubling of the number of degrees of freedom with respect to the real case, which affects the computation of the relic density and therefore the DM mass $M$. In the following we show that, under some conditions, the relic density for $(\bol{1}, \bol{n}, \epsilon)$ can be obtained by simply scaling results for $(\bol{1}, \bol{n}, 0)$ appeared in the literature.

We start by expressing a complex scalar multiplet in terms of its real components, and a Dirac fermion as two degenerate Majorana states with opposite parity under charge conjugation:
\begin{align}
\X = \frac{\X_1 + i \X_2}{\sqrt{2}}
&
\text{ for scalar } \X,
&
\begin{cases}
\X &= \dfrac{\X_1 + \X_2}{\sqrt{2}}
\\
\X^\cc &= \dfrac{\X_1 - \X_2}{\sqrt{2}} \rule{0mm}{8mm}
\end{cases}
&
\text{ for fermion } \X.
\end{align}
In this new basis, and in the absence of quartic couplings in $V(\X, H)$ for scalar MDM, we get two separate Lagrangians for $\X_1$ and $\X_2$ that are bilinear in these fields, and in fact we can define a global $\ZZ_2^{(1)} \times \ZZ_2^{(2)}$ symmetry acting on $(\X_1, \X_2)$ as
\begin{align}
\label{2Z2}
\ZZ_2^{(1)}: (\X_1, \X_2) \to (- \X_1, \X_2) \ ,
&&&
\ZZ_2^{(2)}: (\X_1, \X_2) \to (\X_1, - \X_2) \ .
\end{align}
If we consider now the main annihilation mode of this candidate, \ie $2 \to 2$ DM annihilations into SM vectors $V$, this symmetry tells us that the only possibile annihilation channels are $\X_1 \X_1 \to V V$ and $\X_2 \X_2 \to V V$, since $\X_1 \X_2 \to V V$ has initial and final states with different parity. This means that, at tree level, the two sectors completely decouple.

This is not the whole story, however. As well known, Sommerfeld enhancement provides an important non-perturbative correction relevant in the non-relativistic regime, and thus must be included. Since all relevant diagrams are of ladder type, we have again that if a process is initiated \eg by $\X_i \X_i$ then no $\X_{j \neq i}$ particles appear in the diagram. Therefore Sommerfeld enhancement respects the complete factorization of $\X_1$ and $\X_2$. As a consequence, the computation of the relic density of $\X_1$ is completely independent from the computation of the relic density of $\X_2$. Moreover, these two states having same mass and same gauge interactions, they must have the same relic density and therefore the relic density of $\X$ is twice that of a single $\X_i$.

\begin{figure}[t]
\centering
\includegraphics[width=0.48\textwidth]{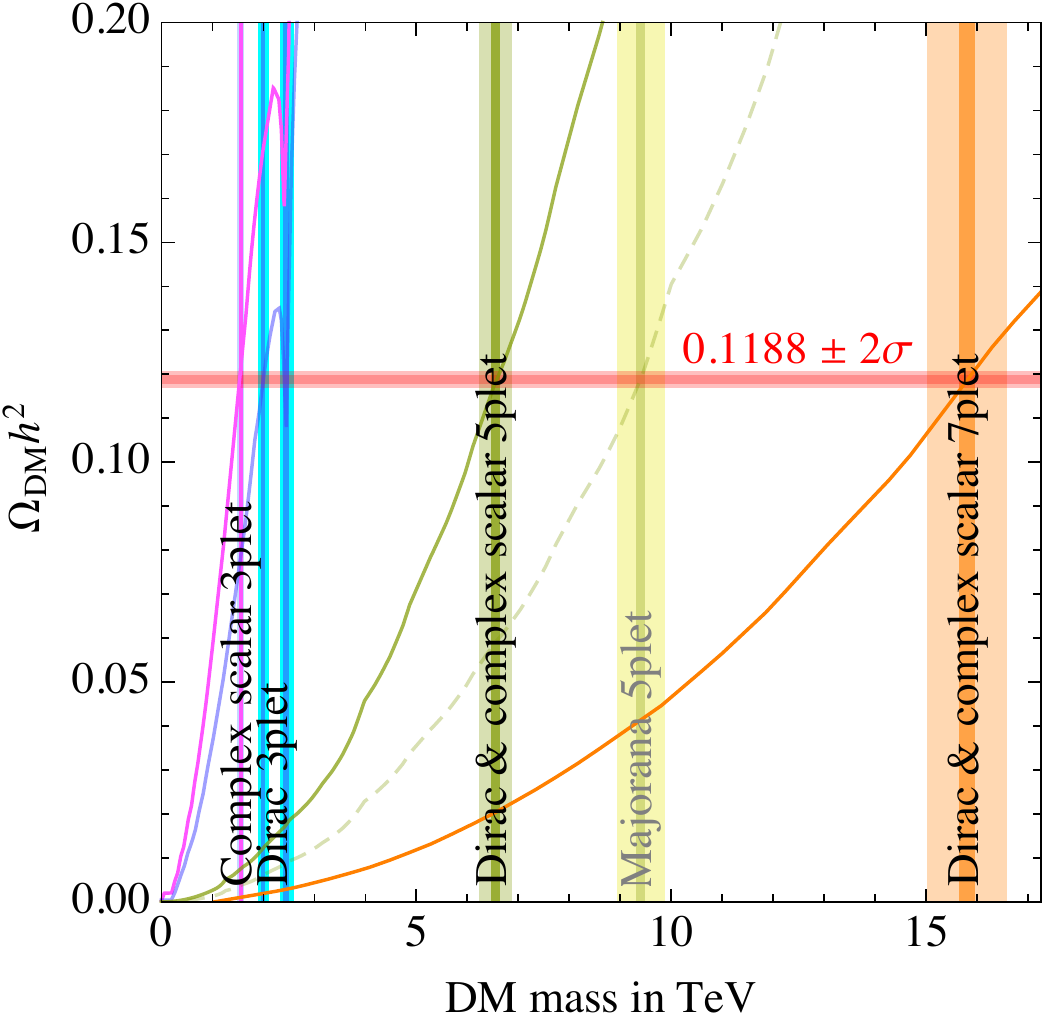}
\includegraphics[width=0.4985\textwidth]{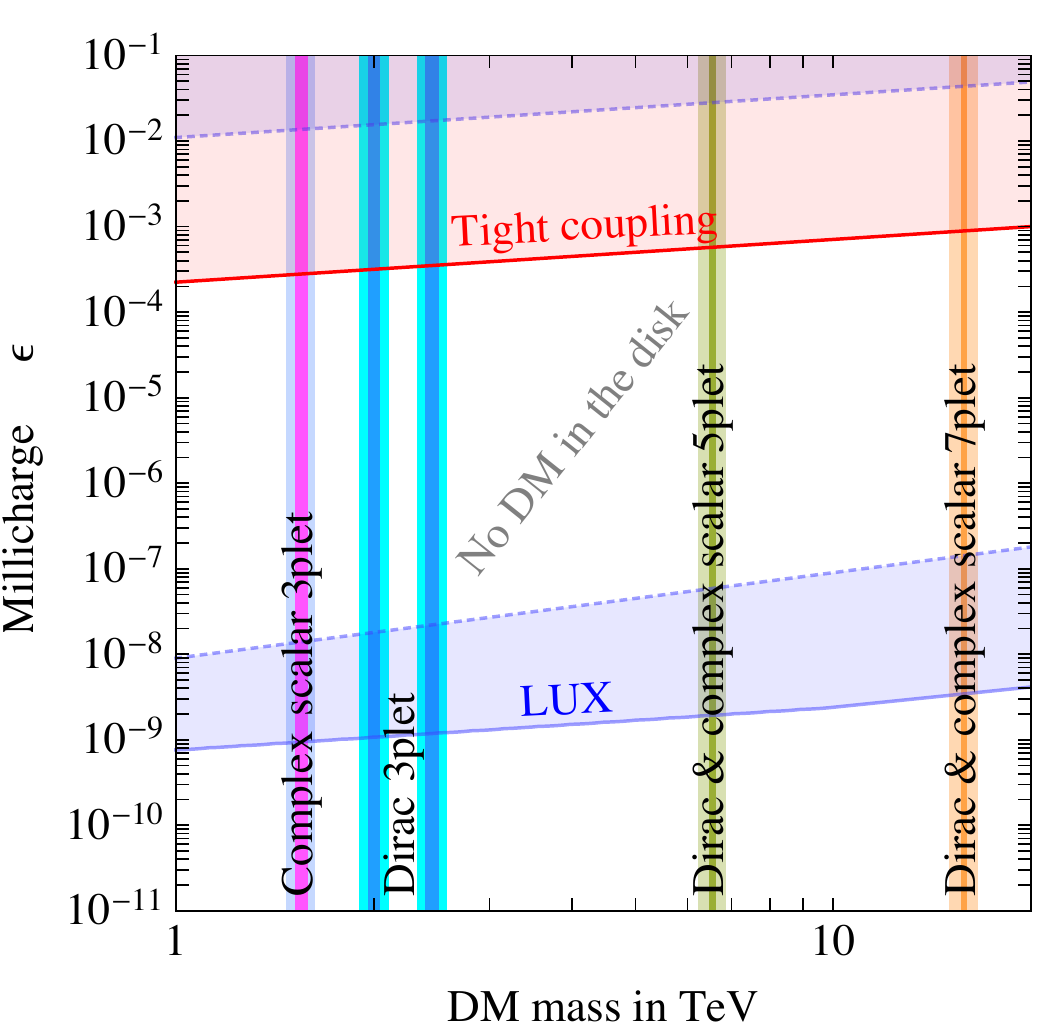}
\caption{\label{fig:Masses}\em
{\bf Left:} Thermal relic abundance of a complex scalar triplet and eptaplet and a Dirac triplet and quintuplet, indicated as solid lines. Confrontation with the measurement by \PLANCK, indicated here as a double horizontal red band (inner for $1 \sigma$ uncertainty, outer for $2 \sigma$), determines the DM mass $M$ in each case. Uncertainties on $M$ are indicated by a double vertical band: the inner, darker band reflects the $2 \sigma$ uncertainty on \PLANCK's measurement, while the outer, lighter band shows the theoretical uncertainty estimated as $\pm 5\%$ of the DM mass. The relic density line for the Dirac triplet crosses the DM abundance band twice, thus there are two allowed values for its mass. We assume the complex scalar quintuplet (eptaplet) has the same mass as the Dirac quintuplet (eptaplet), as happens for real scalar and Majorana quintuplets. The thermal relic abundance of a Majorana quintuplet (dashed line), together with its mass, is shown for use in the next section.
{\bf Right:} Constraints on the DM millicharge $\epsilon$ as a function of the DM mass. The \LUX~bound does not apply in the region of parameter space where no DM particles populate the galactic disk.}
\end{figure}

\Fig{fig:Masses} shows the Sommerfeld-corrected DM relic density as a function of the DM mass for a complex scalar triplet and eptaplet and a Dirac triplet and quintuplet (solid lines). These functions are taken to be twice the value of the relic density of a real scalar triplet~\cite{Cirelli:2007xd}, a real scalar eptaplet in the approximate $\SUd$-symmetric limit~\cite{Garcia-Cely:2015quu}, and a Majorana triplet~\cite{Cirelli:2007xd} and quintuplet~\cite{Cirelli:2015bda}, respectively (the relic density of the Majorana quintuplet is also shown as a dashed line). Since the real scalar quintuplet was found in Ref.~\cite{Cirelli:2007xd} to have the same mass of the Majorana quintuplet, we assume the same holds also for the complex case for both the quintuplet and the eptaplet. While candidates with larger $\bol{n}$ can be perfectly viable, we only consider here $\SUd$ triplets, quintuplets and eptaplets as careful computations of the relic abundance are available in the literature only for these candidates. The horizontal red strips in \Fig{fig:Masses} show \PLANCK's measurement of DM density~\cite{Ade:2015xua},
\beq
\Omega_\text{DM} h^2 = 0.1188 \pm 0.0010 \ ,
\eeq
at the $1 \sigma$ (inner strip) and $2 \sigma$ (outer strip) CL. The DM mass for each case is determined by the crossing of the relevant solid line and the red strips (notice the relic density line for the Dirac triplet crosses the DM abundance band twice, thus there are two allowed values for its mass). This interval in DM masses is indicated with a vertical band whose width is given by the $2 \sigma$ uncertainty of Planck's result. A larger uncertainty, of order $5\%$ of the total result, comes however from the theoretical determination of the cross sections~\cite{Cirelli:2015bda}. This translates into an uncertainty on the determination of the DM mass shown as a lighter vertical band for each case. Considering the latter uncertainty the DM has mass
\begin{align}
1.55 \pm 0.08 \text{ TeV} && \text{complex scalar triplet},
\\
2.00 \pm 0.10 \text{ or } 2.45 \pm 0.12 \text{ TeV} && \text{Dirac triplet},
\\
6.55 \pm 0.33 \text{ TeV} && \text{complex scalar and Dirac quintuplet},
\\
15.8 \pm 0.79 \text{ TeV} && \text{complex scalar and Dirac eptaplet}.
\end{align}
Also shown in \Fig{fig:Masses} is the mass for a Majorana quintuplet (which is determined by its relic density, shown as a dashed line). This information will be useful for our study of this candidate in the next section.

For scalar $\X$, quartic couplings such as $\X^\dagger t^a_\X \X \, H^\dagger t^a_H H$ can break the symmetry in \Eq{2Z2} and thus affect the above scaling argument. In this case, the annihilation cross section will be in general larger than that, discussed above, due solely to DM couplings to gauge bosons. Therefore, in order to fit the observed DM abundance, the DM mass must be larger with respect to the case of DM with only gauge interactions. The values of $M$ given above and shown in \Fig{fig:Masses} can thus be thought of lower bounds on the true value of the DM mass in presence of quartic couplings. See \eg Ref.~\cite{Hambye:2009pw} for a dedicated analysis on the effect of these couplings.

Once the mass of the $(\bol{1}, \bol{n}, \epsilon)$ multiplet is known, the phenomenology of these candidates is univocally determined (up to free terms in $V(\X, H)$ for scalar multiplets). In particular, the most stringent constraints on electroweak multiplets come from indirect DM searches. The bounds from gamma-ray line searches are particularly relevant due to the Sommerfeld-enhanced annihilation cross section into gauge bosons.

The phenomenological advantage of millicharged MDM candidates is that, since the DM particle and its antiparticle are distinct, the annihilation probability is half that of a self-conjugated DM candidate with the same quantum numbers and mass. Therefore, all bounds on the annihilation cross section are a factor of $2$ less stringent. However, the DM mass for these candidates is in general lower than for their self-conjugated version, and this may be a drawback for the following reason. Ideally, bounds on annihilation cross sections scale with the inverse of the DM number density squared, and thus with $(\rho / \mDM)^{-2}$ with $\rho$ the assumed DM energy density. Therefore, lighter DM candidates are ideally more constrained. However, realistic bounds depend on the experimental resolution, which is particularly relevant for gamma-ray line searches given that the expected signal is a very narrow spectral feature. A finite and energy-dependent resolution leads in general to an uneven sensitivity on the DM mass. Moreover, the theoretical dependence of the annihilation cross section on $\mDM$ may be very irregular, especially in presence of non-perturbative effects (see \eg Fig.~7 of Ref.~\cite{Cirelli:2015bda}). Therefore, it is difficult to predict whether a lighter DM particle is more or less constrained than a somehow heavier particle. For this reason, constraints on millicharged MDM candidates must be checked case by case.

Bounds on some of the candidates considered above can be determined by properly rescaling existing bounds on self-conjugated multiplets with the same quantum numbers. Constraints on a (supersymmetric Wino) Majorana triplet, on the MDM Majorana quintuplet, and on the real scalar eptaplet can be found in Refs.~\cite{Hryczuk:2014hpa, Cohen:2013ama, Chun:2015mka, Bhattacherjee:2014dya, Fan:2013faa}, \cite{Cirelli:2015bda, Garcia-Cely:2015dda, Cirelli:2007xd}, and~\cite{Garcia-Cely:2015quu}, respectively. We do not have enough information on the scalar triplet and fermion eptaplet to determine bounds on these candidates.

Interestingly, the Dirac triplet with $\mDM = 2.00$ TeV is allowed by gamma-ray searches even with the most aggressive choices of DM profile made in Fig.~12 of Ref.~\cite{Hryczuk:2014hpa}. In the assumption of a cuspy profile, forthcoming experiments like \CTA~\cite{Consortium:2010bc} will be able to probe this candidate. The situation of the Dirac triplet with $\mDM = 2.45$ TeV is closer to (although worse than) that of the Majorana triplet with mass $3.1$ TeV~\cite{Cohen:2013ama}, which is already excluded by bounds assuming cuspy profiles while allowed when choosing a cored profile. The $6.55$ TeV Dirac quintuplet is in the same situation as the Majorana quintuplet, whose mass is given in \Eq{quintupletmass}, \ie it is badly excluded with the choice of a cuspy profile, while it is still viable if a cored profile is considered (see \eg Fig.~7 of Ref.~\cite{Cirelli:2015bda}). The complex scalar eptaplet, while excluded for a cuspy Einasto profile, may be either excluded or allowed for a cored Isothermal profile, depending on the precise value of its mass within the $5\%$ uncertainty reported above and shown in \Fig{fig:Masses}; notice however that our calculation of mass and constraints for this candidate rely on the computations carried out in~\Ref{Garcia-Cely:2015quu} in the approximate limit of exact $\SUd$ symmetry.

We have only discussed here bounds from gamma-ray line searches, which are the most constraining, as mentioned above, when a cuspy profile is chosen. Other bounds, that a rough evaluation reveals not to exclude these candidates at present, may become relevant in the near future (see \eg Refs.~\cite{Cirelli:2015bda, Hryczuk:2014hpa, Cirelli:2014dsa}). The most entertaining possibility is to probe MDM with a future $100$ TeV proton-proton collider~\cite{Cirelli:2014dsa}, but the hope is to find other evidences for it well before that.

\section{Decaying quintuplet MDM}
\label{Decaying MDM}

In this section we study the possibility that the MDM setup has a generic cutoff $\Lambda$. We consider here the `standard' MDM scenario with $Y = 0$, thus the only viable candidate (as discussed above) is the fermionic $\SUd$ quintuplet. The main effect of lowering the cutoff from the Planck scale is that DM stability on cosmological timescales is spoiled. In fact, dimension-$6$ operators can break the $\ZZ_2$ symmetry protecting DM from decay~\cite{Cirelli:2009uv}, so that for a low-enough cutoff we can expect to observe the signature of DM decays in cosmic-ray spectra. We thus perform a thorough analysis of the gamma-ray spectrum produced in DM decays and use the \FERMI~\cite{Ackermann:2014usa} and \HESS~\cite{Abramowski:2013ax} data to set bounds on $\Lambda$.

\subsection{Relevant Lagrangian and decay modes}
We represent the fermionic $\SUd$ quintuplet as a Dirac four-spinor $\X$ with only right-handed components, so that $P_\RH \X = \X$ and $P_\LH \X = 0$, where $P_\RH$ and $P_\LH$ are the right and left projectors, respectively. Notice that the quintuplet is a real representation of $SU(2)$ and therefore the neutral component of $\X$ is a Majorana fermion. DM decays are induced at dimension $6$ by the two operators $\overline{\X} L H H H^\dagger$, $\overline{\X} \sigma_{\mu \nu} L W^{\mu \nu} H$, and their hermitian conjugates. We are therefore interested in the following Lagrangian:
\beq
\Lag = \Lag_\text{SM} + i \overline{\X} \slashed{D} \X + \left( -\frac{\mDM}{2} \overline{\X^\cc} \X + \frac{c_1^a}{\Lambda^2} \overline{\X} L^a H H H^\dagger + 
\frac{c_2^a}{\Lambda^2} \overline{\X} \sigma_{\mu \nu} L^a W^{\mu \nu} H + \text{h.c.} \right) ,
\eeq
where $a = e, \mu, \tau$ is a lepton flavor index and $\sigma^{\mu \nu} \equiv \frac{i}{2} [\gamma^\mu, \gamma^\nu]$. We neglect dimension-$5$ and all other dimension-$6$ operators, since they do not contribute to DM decays. DM annihilations are of course dominated by $\X$'s renormalizable gauge couplings, as the contribution of non-renormalizable operators is suppressed by powers of $\mDM / \Lambda$. To show how the multiplet components of the various fields contract, we represent $\X$ as a rank-$4$ completely symmetric tensor in the anti-fundamental representation of $\SUd$, $\X^{ijkl}$ with $i, j, k, l = 1, 2$ (see \eg Appendix B of Ref.~\cite{DiLuzio:2015oha} for more details). The $W$-boson multiplet is also written as a symmetric rank-$2$ tensor, while the lepton doublet $L$ and the Higgs doublet $H$ are represented by rank-$1$ tensors in the fundamental representation. Indices are raised and lowered with the completely antisymmetric $\SUd$-invariant tensor $\epsilon$, with $\epsilon^{12} = - \epsilon_{12} = 1$. Making the $\SUd$ indices explicit we get
\begin{multline}
\Lag = \Lag_\text{SM} + i \overline{\X}_{i j k l} \slashed{D}^{i j k l}_{i' j' k' l'} \X^{i' j' k' l'}
- \frac{\mDM}{2} \left( \overline{\X^\cc}^{i j k l} \X^{i' j' k' l'} \epsilon_{i i'} \epsilon_{j j'} \epsilon_{k k'} \epsilon_{l l'} + \text{h.c.} \right)
\\
+ \left( \frac{c_1^a}{\Lambda^2} \, \overline{\X}_{i j k l} L^a_{i'} H_{j'} H_{k'} H^{\dagger l} \epsilon^{i i'} \epsilon^{j j'} \epsilon^{k k'}
+ \frac{c_2^a}{\Lambda^2} \, \overline{\X}_{i j k l} \sigma_{\mu \nu} L^a_{i'} W^{\mu \nu}_{j' k'} H_{l'} \epsilon^{i i'} \epsilon^{j j'} \epsilon^{k k'} \epsilon^{l l'} + \text{h.c.} \right) .
\end{multline}
The fields can be rewritten as
\begin{align}
\begin{array}{l}
\X^{1111} = \X^{-2}_\RH \\
\X^{1112} = \tfrac{1}{\sqrt{4}} \X^{-1}_\RH \rule{0mm}{5mm} \\
\X^{1122} = \tfrac{1}{\sqrt{6}} \X^0_\RH \rule{0mm}{5mm} \\
\X^{1222} = \tfrac{1}{\sqrt{4}} \X^{+1}_\RH \rule{0mm}{5mm} \\
\X^{2222} = \X^{+2}_\RH \rule{0mm}{5mm}
\end{array}
&&
\begin{array}{l}
W_{11} = + \tfrac{1}{\sqrt{2}} W^+ \\
W_{12} = - \tfrac{1}{2} W^3 \rule{0mm}{5mm} \\
W_{22} = - \tfrac{1}{\sqrt{2}} W^- \rule{0mm}{5mm}
\end{array} 
&&
\begin{array}{l}
H_{1} = \phi^+ \\
H_{2} = \frac{1}{\sqrt{2}} (h + v + i \phi^0) \rule{0mm}{5mm} \\
L_1 = \nu_\LH \rule{0mm}{5mm} \\
L_2 = \ell_\LH \rule{0mm}{5mm}
\end{array} 
\end{align}
where the $\X_\RH^{t^3}$ are Dirac spinors with only right-handed components. The DM candidate is the self-conjugated neutral component $\X^0_\RH$, which from now on we will denote $\DM$ for simplicity. As can be seen from \Fig{fig:Masses}, the DM mass is fixed by its relic abundance to be
\beq\label{quintupletmass}
\mDM = 9.4 \pm 0.47 \text{ TeV} \ .
\eeq

We study the effect of the two dimension-$6$ operators separately. Detailed analytic formulas for the matrix elements and the differential and total decay rates for the relevant processes can be found in Appendix \ref{app:rates}. Since the DM is much heavier than all SM particles it decays to, we neglect all final state particle masses in the calculations.

The first operator, $\overline{\X} L H H H^\dagger$, induces $\nu_\LH$-$\DM$ mixing and the following $2$, $3$, and $4$-body DM decays:
\beq\label{decaymodes1}
\begin{split}
\DM \to & \ \ell W_\LH^+, \nu Z_\LH, \nu h,
\\
\DM \to & \ \ell W_\LH^+ h, \nu W_\LH^+ W_\LH^-, \nu Z_\LH Z_\LH, \nu Z_\LH h, \nu h h,
\\
\DM \to & \ \ell W_\LH^+ W_\LH^+ W_\LH^-, \ell W_\LH^+ Z_\LH Z_\LH, \ell W_\LH^+ h h,
\\
& \ \nu Z_\LH Z_\LH Z_\LH, \nu Z_\LH W_\LH^+ W_\LH^-, \nu Z_\LH Z_\LH h, \nu W_\LH^+ W_\LH^- h, \nu Z_\LH h h, \nu h h h,
\end{split}
\eeq
where all gauge bosons are longitudinal and charged leptons and neutrinos are left-handed (the polarization of final state particles is an important ingredient entering the code~\cite{Cirelli:2010xx}, that we use to compute the gamma-ray flux from DM decays). Notice that the squared amplitude for decays into final states with many Higgs fields are enhanced by powers of $(\mDM / v)^2 \approx 10^3$: in fact, adding a Higgs field to the final state removes a factor of $v$ from the Lagrangian coefficient, that is replaced in the decay amplitude by a factor of $\mDM$. By the Equivalence Theorem, the same holds for decays into many longitudinal gauge bosons as well. Therefore, despite the phase-space suppression, decays into many particles can be favored over $2$-body decays. We check explicitly that this is the case by computing both $\DM \to \nu h$, $\nu h h$, and $\nu h h h$ decay rates. The remaining decay rates are computed with the Equivalence Theorem. All our analytic results can be found in Appendix \ref{app:rates}. Our analytic computation of the $4$-body phase space (approximating all final states as massless) is described in Appendix \ref{app:phasespace}.

The second operator, $\overline{\X} \sigma_{\mu \nu} L W^{\mu \nu} H$, induces the following $2$ and $3$-body decays:
\beq\label{decaymodes2}
\begin{split}
\DM \to & \ \ell W_\TT^+, \nu Z_\TT, \nu \gamma,
\\
\DM \to & \ \ell W_\TT^+ Z_\LH, \ell W_\TT^+ h, \ell Z_\TT W_\LH^+, \ell \gamma W_\LH^+, \nu Z_\TT h, \nu Z_\TT Z_\LH, \nu \gamma Z_\LH, \nu \gamma h, \nu W_\TT^- W_\LH^+,
\end{split}
\eeq
where one gauge boson is always transverse, while the other, if present, is longitudinal, and the charged lepton or neutrino is left-handed. Contrary to the previous case, $4$-body decays do not receive the $(\mDM / v)^2$ enhancement factor with respect to the $3$-body modes, and thus can be neglected. $2$-body decays are also suppressed with respect to the $3$-body channels, but they deserve special attention since they produce very narrow features in the gamma-ray spectrum. These peaks, appearing at the very end of the produced photon spectrum (\ie at energies equal to half the DM mass), are due to photons produced by the monochromatic decay products of $2$-body decays, and may be visible on top of the continuum produced by $3$-body decays. In Appendix \ref{app:rates} we compute explicitly the $\DM \to \ell W, \nu Z, \nu \gamma, \ell W h, \nu Z h, \nu \gamma h$ decay rates, and apply the Equivalence Theorem to compute the remaining rates. To avoid the shortcomings of the Theorem (see \eg footnote 7 of~\cite{Wulzer:2013mza}) we checked the result by also performing the computation in the Equivalent gauge~\cite{Wulzer:2013mza}.

\subsection{Gamma-ray fluxes from DM decay}
The strongest limit on models of decaying DM is arguably set by observations of gamma rays. For this reason we focus here on production of secondary gamma rays, and compare the model expectation with \FERMI~data to obtain a bound on the relevant parameter $c_{1, 2}^a / \Lambda^2$. Moreover,
the photon flux does not suffer from the same astrophysical (\eg diffusive) uncertainties as charged particles, thus our analysis is quite reliable and does not depend much on the modeling of the cosmic environment. Other relevant constraints may be set for instance by looking at the sum of electron and positron fluxes up to $1$ TeV measured by \FERMI~and more recently by {\sf AMS-02}, and above that energy by \HESS~and \MAGIC, at the positron fraction and the antiproton flux measured by both {\sf AMS-02}~and \PAMELA, and at the neutrino flux in {\sf ICECUBE} and {\sf ANTARES}. For a recent review of the status of indirect DM searches see \eg Ref.~\cite{Cirelli:2015gux}.

\subsubsection{Production rate of stable SM particles in DM decays}
Upon decay of the DM particle, $\DM \to f_1 \dots f_n$, the primary decay products $f$ undergo a series of processes such as decay and radiative processes (hadronization, showering, ...) which generate a set of stable SM particles $\alpha = e^\pm, \bar p, \gamma, \dots$. The production rate of each stable state $\alpha$ at the source per single DM decay is
\beq\label{alpha spectrum}
\frac{\ud R^\text{s}_\alpha}{\ud E_\alpha}(E_\alpha) = \sum_f \int \ud E_f \, \frac{\ud \Gamma}{\ud E_f}(E_f) \frac{\ud N^{f}_\alpha}{\ud E_\alpha}(E_f, E_\alpha) \ ,
\eeq
where $\ud N^{f}_\alpha / \ud E_\alpha$ is the spectrum arising solely from the primary $f$ with energy $E_f$, and $\ud \Gamma / \ud E_f$ is the DM decay rate into $f$ summed over all decay channels which include $f$ in the final state.

While propagating away from the source, these stable particles can interact with the cosmic environment thus modifying their spectrum in a position-dependent way. For instance, the photon flux at Earth gets a contribution from the prompt emission in \Eq{alpha spectrum} with $\alpha = \gamma$, and a contribution from low-energy background photons (\eg from the CMB or the interstellar photon field) being up-scattered by $e^\pm$ from DM decays; in the latter case, the $\alpha = e^\pm$ rate at the source in \Eq{alpha spectrum} must be convolved with the probability of undergoing inverse-Compton (IC) processes with the inhomogeneous photon field (see e.g. Refs.~\cite{Blumenthal:1970gc, Cirelli:2009vg}). The so-modified rate, which we call $\ud R_\alpha / \ud E_\alpha$, depends \eg on the distance $r$ from the Galactic Center (GC) for decays within our galaxy, or from the redshift $z$ for extragalactic decays. We compute $\ud R_\alpha / \ud E_\alpha$ from the production rate at the source $\ud R^\text{s}_\alpha / \ud E_\alpha$ following Ref.~\cite{Cirelli:2010xx}.

We adapt the spectra per single primary $\ud N^{f}_\alpha / \ud E_\alpha$ from Ref.~\cite{Cirelli:2010xx}, with the following cautions. The primary spectra given in Ref.~\cite{Cirelli:2010xx} are meant for DM decays to particle-antiparticle pairs $\DM \to f \overline{f}$, so that the primary energies $E_f$ are not parameters that can be varied but are instead fixed to half the input DM mass, call it $M_\text{PPPC}$. The latter is a parameter whose value is possible to vary, and therefore in using the primary spectra from Ref.~\cite{Cirelli:2010xx} we adopt the prescription $M_\text{PPPC} \to 2 E_f$. One has also to take into account the fact that the primary spectra given in Ref.~\cite{Cirelli:2010xx} include the spectrum generated by the primary antiparticle $\overline{f}$, besides that due to $f$. However, in the assumption of CP invariance, the rate for the decay $\DM \to f_1 \dots f_n$ equals the rate for $\DM \to \overline{f_1} \dots \overline{f_n}$ (notice that $\DM$ is a Majorana fermion). Therefore, when summing the two rates (as part of the sum over all decay channels), we will have in \Eq{alpha spectrum} $\ud N^{f}_\alpha / \ud E_\alpha + \ud N^{\overline{f}}_\alpha / \ud E_\alpha$. We use the spectra of Ref.~\cite{Cirelli:2010xx} in place of this sum. Consequently, the only channels that remain to be summed are those that are non mutually conjugated. In practice \Eq{alpha spectrum} can be operatively written as
\beq
\frac{\ud R^\text{s}_\alpha}{\ud E_\alpha}(E_\alpha) =
\sum_c \sum_f \int \ud E_f \, \frac{\ud \Gamma_c}{\ud E_f}(E_f)
\hspace{-0.65cm}
\underbrace{
\left[ \frac{\ud N^{f}_\alpha}{\ud E_\alpha} + \frac{\ud N^{\overline{f}}_\alpha}{\ud E_\alpha} \right]
}_{
\displaystyle \left. \frac{\ud N^{\text{PPPC}, f \overline{f}}_\alpha}{\ud E_\alpha} \right|_{M_\text{PPPC} = 2 E_f}
} \ ,
\eeq
where $c$ enumerates all non mutually-conjugated decay channels.

\subsubsection{Continuum photon emission}
The residual isotropic gamma-ray flux observed by \FERMI~\cite{Ackermann:2014usa} extends from $100$ MeV up to $820$ GeV. The origin of this isotropic emissions is not well understood and can be due to different phenomena such as unresolved sources or truly diffusive processes.
 
DM decays can contribute to this isotropic flux, with two components:
\textit{i)} galactic (Gal) residual flux due to DM decays within the Milky Way halo, and
\textit{ii)} extragalactic (ExGal) flux due to cosmological DM decays integrated over redshift.
The latter is of course isotropic, while the former is not, however its minimum constitutes an irreducible contribution to the isotropic flux. Therefore, the isotropic diffuse gamma-ray flux as measured by \FERMI~is
\beq\label{eq:maingammaflux}
\frac{\ud \Phi_\text{isotropic}}{\ud E_\gamma}
= \frac{\ud \Phi_\text{ExGal}}{\ud E_\gamma} + 4 \pi \min_\Omega \frac{\ud \Phi_\text{Gal}}{\ud E_\gamma \, \ud \Omega} \ .
\eeq
Here we make the reasonable approximation that the minimum of the angular flux in the galaxy is found at the anti-GC, as in Refs.~\cite{Cirelli:2009dv, Cirelli:2012ut, Masina:2012hg}.
This approximation is well justified because, for the decay channels relevant in our analysis, the prompt flux (which is the dominant contribution) follows the angular distribution of DM density, which is of course minimum at the anti-GC. For reasonable sizes of the diffusive halo, moreover, we expect also the IC contribution to approximately follow the angular DM distribution.

The flux from the galactic halo, observed from a given direction and within a solid angle $\ud \Omega$, is in general given by
\beq\label{fluxdec}
\frac{\ud \Phi_\text{Gal}}{\ud E_\gamma \, \ud \Omega} = \frac{r_\odot}{4 \pi} \frac{\rho_\odot}{\mDM} \int_\text{l.o.s.} \frac{\ud s}{r_\odot} \frac{\rho_\text{halo}(r(s, \psi))}{\rho_\odot} \, \frac{\ud R_\gamma}{\ud E_\gamma}(r(s, \psi), E_\gamma) \ ,
\eeq
where $r_\odot = 8.33$ kpc is the Sun's distance from the Galactic Center, $\rho_\odot = 0.3$ GeV/cm$^3$ is the local DM energy density, and $r(s, \psi) = \sqrt{r_\odot^2 + s^2 - 2 r_\odot s \cos \psi}$ is the distance of the annihilation site from the GC, with $s$ parametrizing the distance along the line of sight (l.o.s.) and $\psi$ the angle between the direction of observation and the GC. For the galactic distribution of DM we assume a Navarro-Frenk-White (NFW) profile~\cite{Navarro:1995iw}
\beq
\rho_\text{halo}(r) = \rho_\text{s} \frac{r_\text{s}} r \left(1 + \frac{r}{r_\text{s}} \right)^{-2} \ ,
\eeq
with parameters $\rho_\text{s} = 0.184$ GeV/cm$^3$ and $r_\text{s} = 24.42$ kpc. In any case, given the linear dependence of the photon flux on the DM density for decaying DM (as opposed to the quadratic dependence for annihilating DM), and the fact that we are mainly interested in the anti-GC where all profiles are similar, the final result will bear little dependence on this choice of profile and parameters.

The galactic gamma-ray flux has two main components: \textit{i)} the prompt gamma rays originating from the fragmentation of the primary products of decay, whose spectrum can be obtained by taking $\alpha = \gamma$ in \Eq{alpha spectrum} for all decay channels given above; and \textit{ii)} IC gamma rays, produced by the up-scattering of low-energy photons of CMB, infrared light, and starlight, by energetic $e^\pm$ produced by DM decays. To obtain the IC spectrum we integrate the $\alpha = e^\pm$ rate in \Eq{alpha spectrum} with the IC halo functions given in Refs.~\cite{Cirelli:2010xx, Buch:2015iya}.

The extragalactic flux, integrated over the redshift at emission $z$, is given by
\beq\label{smoothedmap}
\frac{\ud \Phi_\text{ExGal}}{\ud E_\gamma} = \frac{1}{H_0} \frac{\rho_{\text{c}, 0} \, \Omega_\text{DM}}{\mDM}
\int_0^\infty \ud z \, \frac{e^{- \tau(z, E_\gamma)}}{\sqrt{(1 + z)^3 \Omega_\text{M} + \Omega_\Lambda}} \, \frac{\ud R_\gamma}{\ud E'_\gamma}(z, E'_\gamma) \ ,
\eeq
where $H_0$ is today's Hubble expansion rate and $\Omega_\text{M}$, $\Omega_\text{DM}$, $\Omega_\Lambda$ are respectively the matter, DM, and cosmological constant energy density in terms of today's critical density $\rho_{\text{c}, 0}$. $E'_\gamma = (1 + z) E_\gamma$ is the photon energy at emission redshift $z$ so that the same photon is detected on Earth with energy $E_\gamma$. The factor $e^{- \tau(z, E_\gamma)}$ accounts for the absorption of DM-produced gamma rays due to scattering off low-energy background photons, which results in production of energetic $e^\pm$ pairs. The converted energy is in turn redistributed to lower-energy gamma rays via IC scattering off CMB photons. We take into account this effect in our analysis, which is sizable for channels with a pronounced prompt emission (the most relevant cases being $\DM \to \nu \gamma, e W^+, \mu W^+$). We take the optical depth of the Universe $\tau(z, E_\gamma)$ from Ref.~\cite{Cirelli:2010xx}.

As for the galactic flux, the extragalactic spectrum is again the sum of the prompt and IC contributions. The $z$ dependence of the prompt flux is obtained by simply ``redshifting'' $E_\gamma$,
\beq
\frac{\ud R^\text{P}_\gamma}{\ud E'_\gamma}(z, E'_\gamma) = \frac{\ud R^\text{P}_\gamma}{\ud E_\gamma}(0, (1+z) E_\gamma) \ ,
\eeq
while the IC contribution due to $e^\pm$ scattering off the warmer CMB photons at $z > 0$ scales as~\cite{Cirelli:2009dv}
\beq
\frac{\ud R^\text{IC}_\gamma}{\ud E'_\gamma}(z, E'_\gamma) = \frac{1}{1+z} \frac{\ud R^\text{IC}_\gamma}{\ud E_\gamma}(0, E_\gamma) \ .
\eeq

\begin{figure}[t]
\centering
\includegraphics[width=0.49\textwidth]{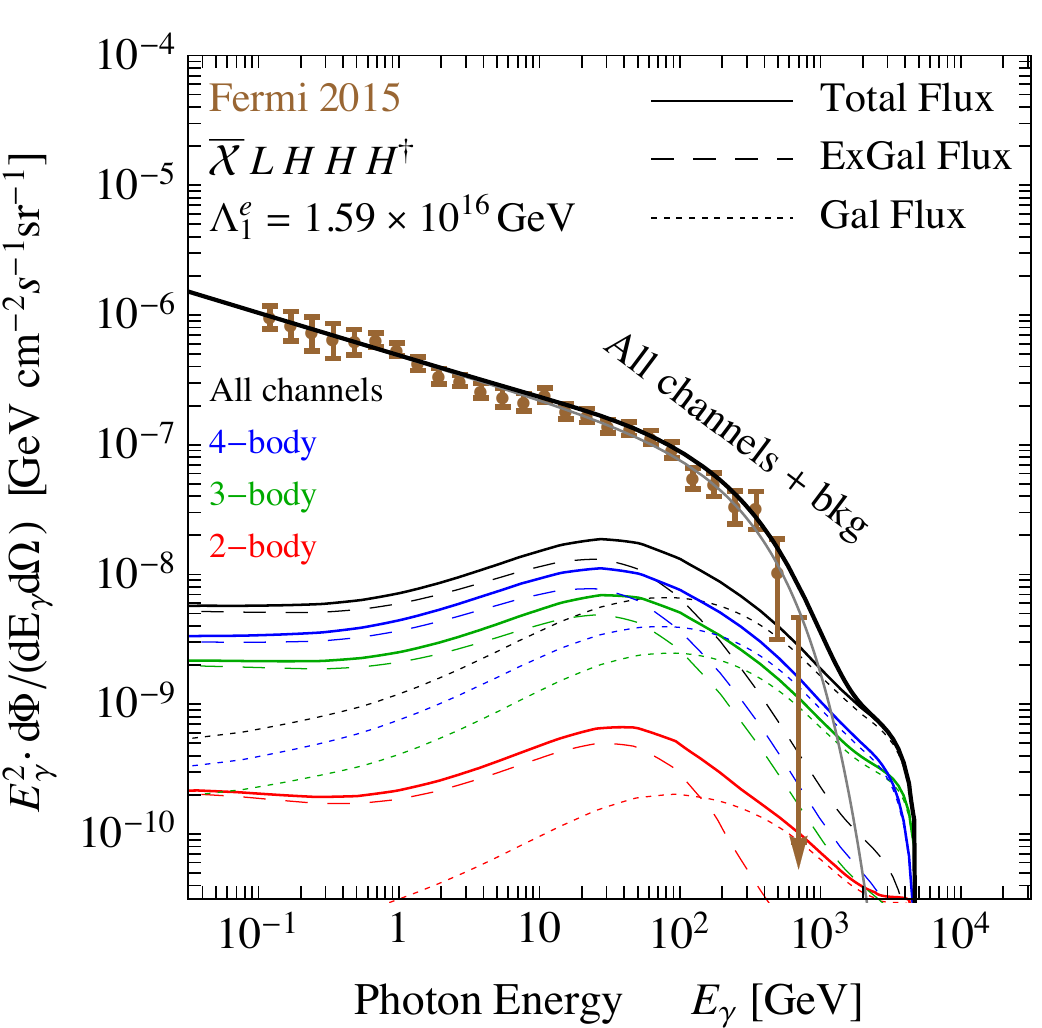}
\includegraphics[width=0.49\textwidth]{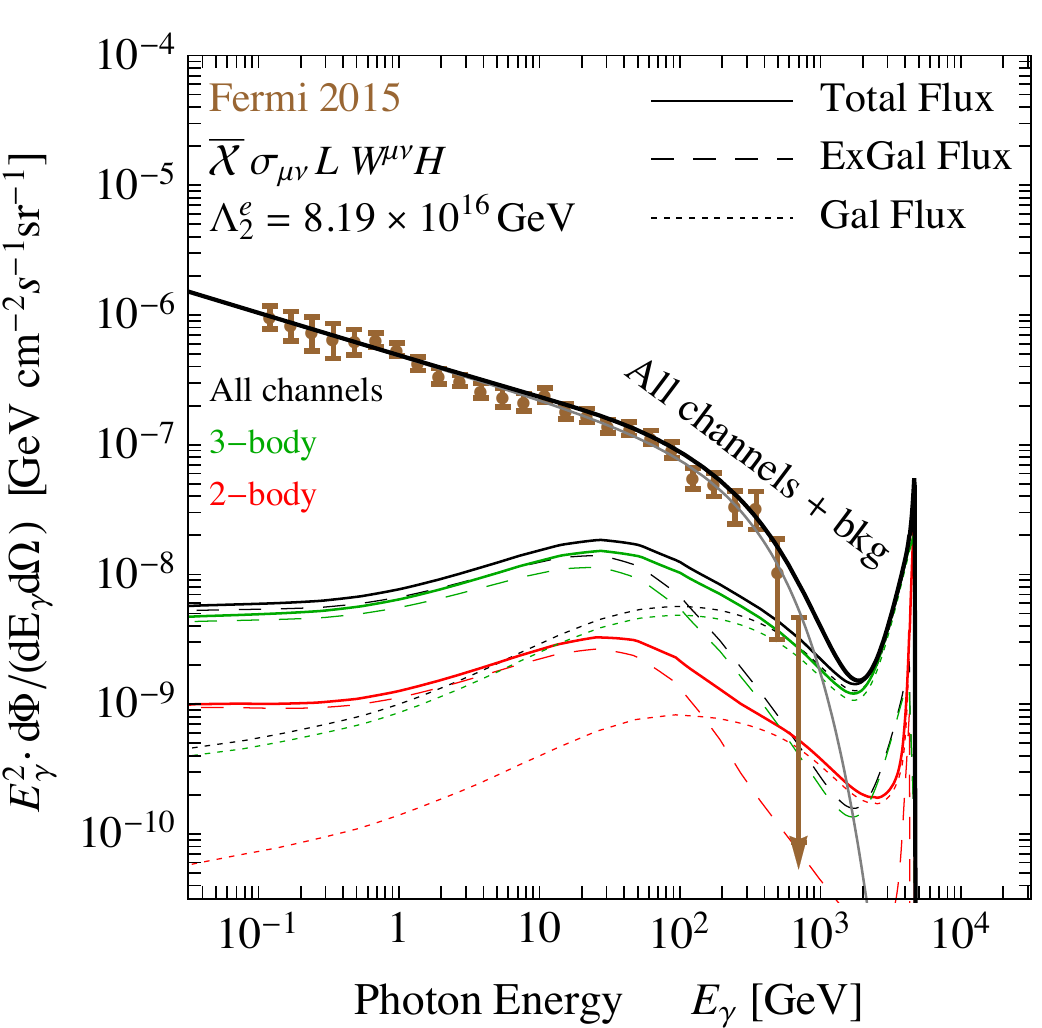}
\caption{\label{fig:quintuplet e}\em
Isotropic gamma-ray flux due to DM decays induced by the operators $(\Lambda_1^a)^{-2} \overline{\X} L^a H H H^\dagger$ ({\bf left}) and $(\Lambda_2^a)^{-2} \overline{\X} \sigma_{\mu \nu} L^a W^{\mu \nu} H$ ({\bf right}), assuming DM coupling to electrons and electron neutrinos ($a = e$). Fluxes from $2$, $3$ and $4$-body decays are separately shown in red, green, and blue, respectively, while the total flux is in black. Dashed lines indicate the extra-galactic component of the flux, dotted lines the galactic flux, and solid lines their sum. \FERMI~data on the diffuse isotropic gamma-ray flux are shown in brown, and the astrophysical background is displayed as a gray line. The thick black line indicates the sum of the total flux from DM decays and the background. The best-fitting value of $\Lambda_{1, 2}^a$, adopted here to normalize the fluxes, is reported in the upper part of the plots.}
\end{figure}

\begin{figure}[t]
\centering
\includegraphics[width=0.49\textwidth]{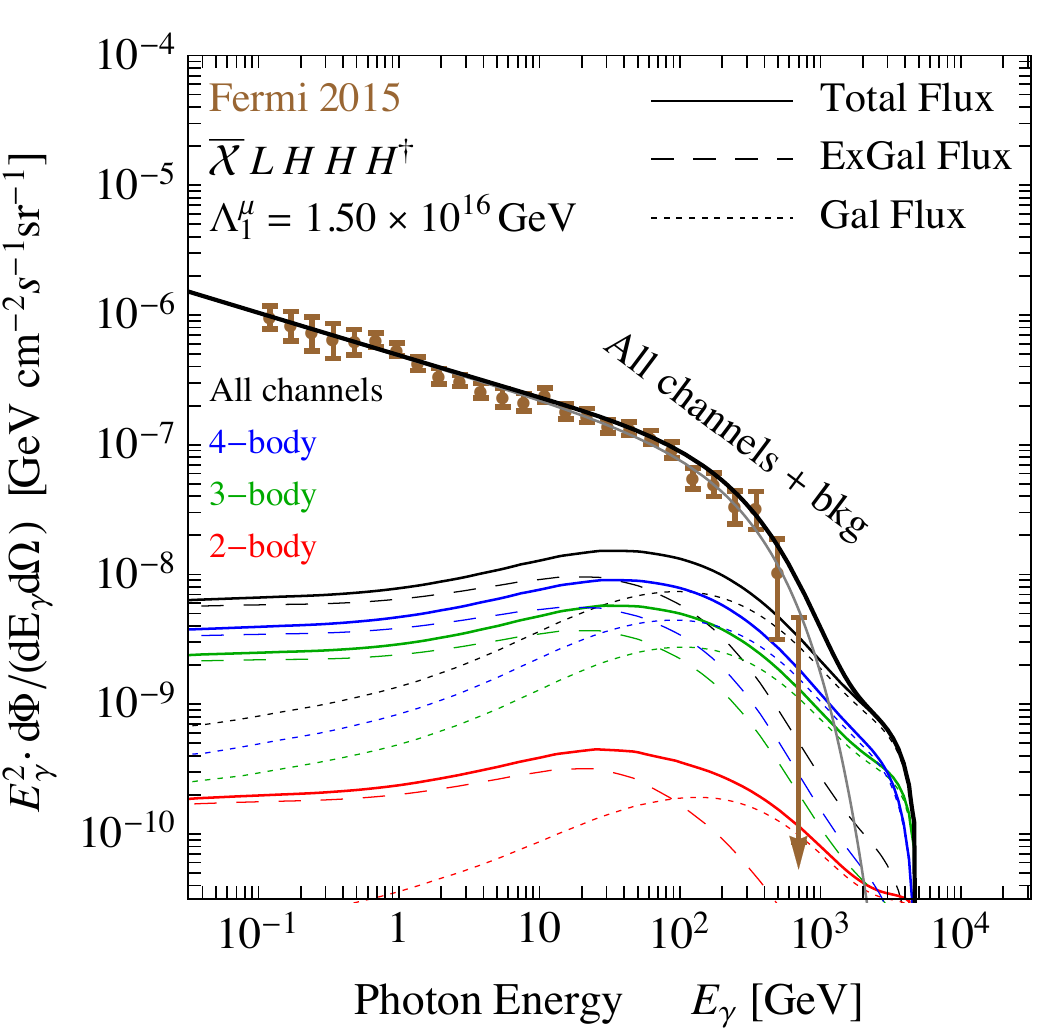}
\includegraphics[width=0.49\textwidth]{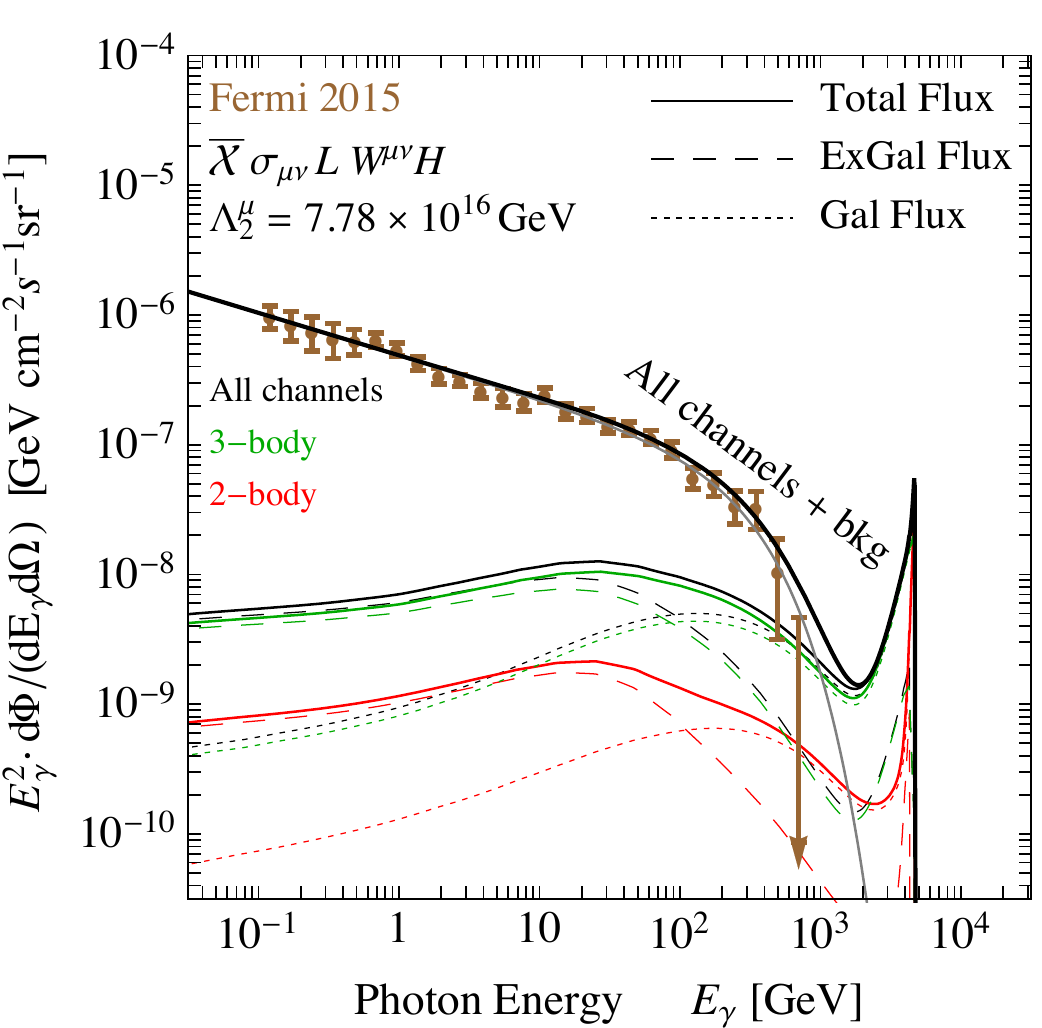}
\caption{\label{fig:quintuplet mu}\em
Same as in \Fig{fig:quintuplet e} but for DM coupling to muons and muon neutrinos ($a = \mu$).}
\end{figure}

\begin{figure}[t]
\centering
\includegraphics[width=0.49\textwidth]{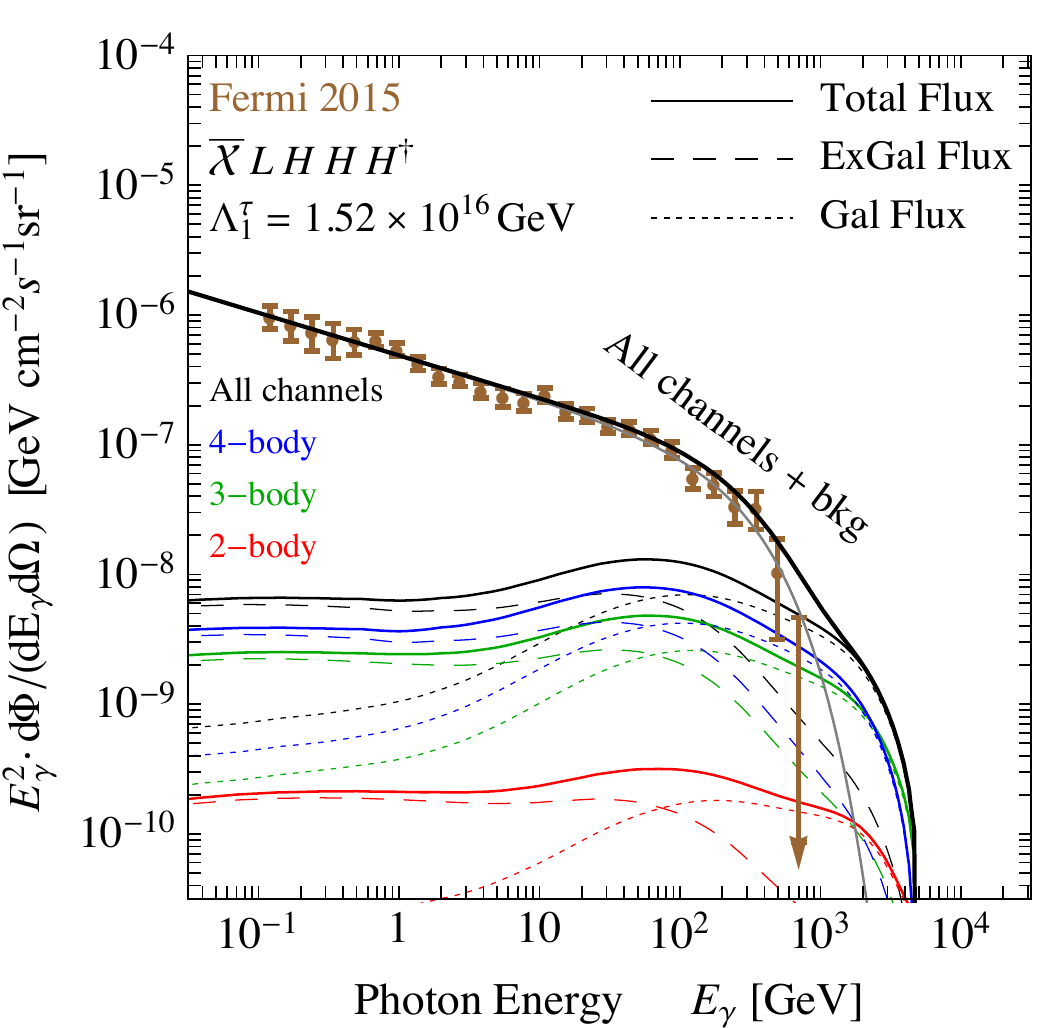}
\includegraphics[width=0.49\textwidth]{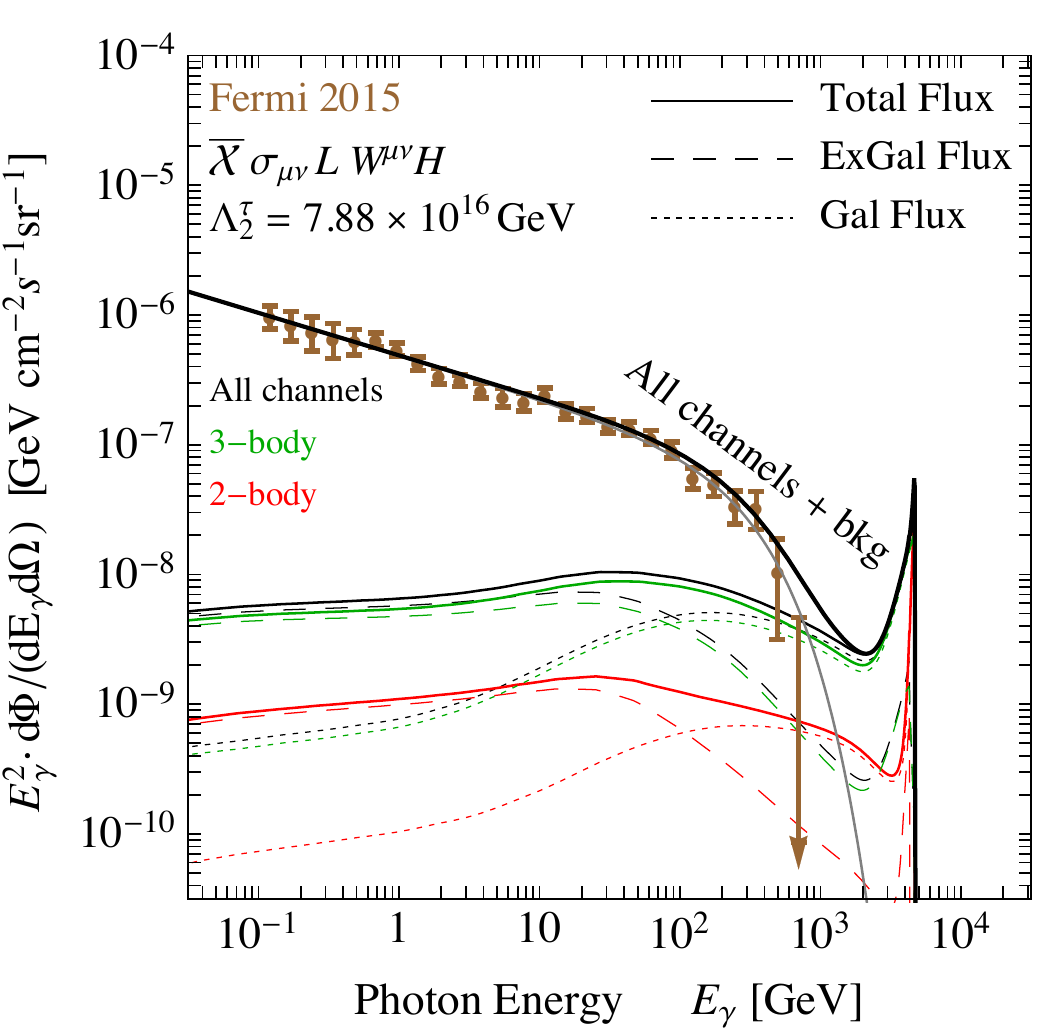}
\caption{\label{fig:quintuplet tau}\em
Same as in \Fig{fig:quintuplet e} but for DM coupling to tau leptons and tau neutrinos ($a = \tau$).}\end{figure}

Figs.~\ref{fig:quintuplet e}, \ref{fig:quintuplet mu}, \ref{fig:quintuplet tau} show the isotropic gamma-ray flux from decaying quintuplet MDM, compared with the flux measured by \FERMI~\cite{Ackermann:2014usa} (brown data points, taken from Table 3 of Ref.~\cite{Ackermann:2014usa}; see also the data table at~\cite{FERMIdata}). Each figure shows the result of our analysis assuming $\DM$ decays only to a specific lepton flavor (\Fig{fig:quintuplet e} for $a = e$, \Fig{fig:quintuplet mu} for $a = \mu$, and \Fig{fig:quintuplet tau} for $a = \tau$). The left panel of each figure shows the flux expected from decays induced solely by the $\frac{c_1^a}{\Lambda^2} \overline{\X} L H H H^\dagger$ operator, while the right panel shows the flux due to decays from $\frac{c_2^a}{\Lambda^2} \overline{\X} \sigma_{\mu \nu} L W^{\mu \nu} H$. In each plot the photon flux is broken into contributions from $2$, $3$ and (only for the first operator) $4$-body decays, respectively in red, green, and blue. Each contribution is in turn separated into its extra-galactic (dashed lines) and galactic (dotted lines) components, with the solid line of same color showing the sum of the two. The thin black lines (again dashed for the extra-galactic component, dotted for the galactic component, and solid for the total), then show the sum of $2$, $3$ and (only for the first operator) $4$-body fluxes. Finally, the solid thick black line shows the sum of the total photon flux from DM decays with the astrophysical background, indicated with a solid gray line.

Clearly the DM signal does not agree in shape with the data, which are instead well fitted by a simple power law with an exponential cutoff. We therefore use this functional form to model the background, adopting the best-fit parameter values from Table 4 of Ref.~\cite{Ackermann:2014usa} and the \FERMI~baseline diffuse galactic emission model (model A of Ref.~\cite{Ackermann:2014usa}). Each plot reports the value of
\begin{align}
\label{Lambdaia}
\Lambda_i^a \equiv \frac{\Lambda}{\sqrt{| c_i^a |}} \ ,
&&
i = 1, 2,
&&
a = e, \mu, \tau,
\end{align}
used to normalize the photon flux from DM decays, which has been determined in each case by performing a minimum-$\chi^2$ analysis of model + background against the \FERMI~data.

As apparent from a comparison of the three figures for each operator, the total DM signal for each operator bears little sensitivity to the lepton flavor $a$, the most noticeable difference being the size of the broad bump peaking between $10$ and $100$ GeV. Since this bump is due to IC processes that populate the high-energy gamma-ray spectrum at the expense of high-energy $e^\pm$, DM decays generating more $e^\pm$ are expected to make it larger. For this reason, the bump is largest for DM coupling to $a = e$, somehow smaller for $a = \mu$, and smallest for $a = \tau$ since $\tau$'s mainly decay into hadrons.

The main difference between the photon fluxes of the two operators is the narrow feature at the very high-energy end of the spectrum for $\overline{\X} \sigma_{\mu \nu} L W^{\mu \nu} H$, which is absent for the other operator. This is due to the prompt photon emission, especially from the $2$-body $\DM \to \nu \gamma$ decay which generates a gamma-ray line at $E_\gamma = \mDM / 2$. $3$-body decays with $\gamma$'s in their final states also contribute to the peak, although with a broader spectrum. We will analyze this feature in more detail in the next section, where we derive a complementary bound coming from line-like searches with the \HESS~telescope.

In deriving a bound on the maximum allowed DM signal, \ie on the minimum value of $\Lambda_i^a$ allowed by data, we adopt two methods.
\begin{itemize}
\item \emph{DM signal only.} This method, which yields a very conservative limit, consists in demanding that the gamma-ray flux from DM decays alone, \ie assuming no background, does not exceed any one of the \FERMI~data points by more than a given significance, which we take to be $3 \sigma$. This option is largely conservative for two main reasons. First, it is quite clear from Figs.~\ref{fig:quintuplet e}, \ref{fig:quintuplet mu}, \ref{fig:quintuplet tau}, that allowing the flux to exceed one data point would result in excesses in nearby data points as well, and therefore the global significance of the exclusion is in principle higher than the chosen significance in one single bin. This is due to the smooth nature of the DM signal (and of the data) in the \FERMI~energy window. Second, the assumption of a negligible background is clearly physically untenable. In fact, the spectral shape of the signal is so different from the data that background is needed in order to obtain a good fit.
\item \emph{DM signal + background.} A more realistic method consists in demanding that the sum of astrophysical background and DM signal does not exceed a chosen level of global significance, which we take to be $3 \sigma$.
\end{itemize}
The values of $\Lambda_i^a$ allowed by \FERMI~data as computed with both methods are summarized in Table~\ref{LambdaBounds}, together with the respective bounds on the DM lifetime $\tau_\text{DM}$. The constraints on $\tau_\text{DM}$ can be compared for a reference with the bounds obtained in Refs.~\cite{Cirelli:2012ut, Ackermann:2012rg} from an earlier \FERMI~data release. The enhanced constraining power of the new data set is apparent. One also needs to bear in mind that, compared to the phenomenological analysis carried out in Ref.~\cite{Cirelli:2012ut}, where only one decay mode is present at a time, the DM candidate considered here features several decay modes some of which contributing negligibly to the gamma-ray emission. For this reason, our bounds on $\tau_\text{DM}$ may appear less strong than naively expected.

\subsubsection{Gamma-ray lines}
As commented above, the gamma-ray flux from DM decays induced by the $\overline{\X} \sigma_{\mu \nu} L W^{\mu \nu} H$ operator displays a sharp feature at energies close to $\mDM / 2$ (see right panels of Figs.~\ref{fig:quintuplet e}, \ref{fig:quintuplet mu}, and \ref{fig:quintuplet tau}). This is due to the presence of decay channels with prompt photon emission, most notably the $\DM \to \nu \gamma$ decay which generates a gamma-ray line at $E_\gamma = \mDM / 2 \approx 5$ TeV. This feature is not constrained by \FERMI~measurements, which only reach up to $820$ GeV. Therefore we compute here a bound from \HESS~gamma-ray line searches~\cite{Abramowski:2013ax}, which extend up to $25$ TeV.

The \HESS~Collaboration performed two separate searches for line-like features in the gamma-ray flux in two sky regions of interest, namely the extragalactic sky and the central galactic halo (CGH) region, the latter defined as a circle of $1^\circ$ radius around the GC, where the Galactic plane is excluded by requiring $|b| > 0.3^\circ$. We compare our gamma-ray flux with \HESS~limits in both sky regions, thus producing two sets of bounds. To take into account the finite experimental resolution we convolve the photon flux with a Gaussian $G(E_\gamma, E)$ centered around $E_\gamma$, where $E$ denotes the energy detected by the instrument. We take the Gaussian function to have resolution $15\%$ of $E_\gamma$~\cite{Abramowski:2013ax, Aisati:2015ova}. We then integrate the signal over each bin in detected energy,
\beq
\int_\text{bin} \ud E \int \ud E_\gamma \, \frac{\ud \Phi}{\ud E_\gamma \, \ud \Omega} G(E_\gamma, E) \ ,
\eeq
and compare the result with the $95\%$ CL limits on the gamma-ray flux in both sky regions shown in Fig.~2 of Ref.~\cite{Abramowski:2013ax}. We neglect IC processes as they only contribute to the continuum gamma-ray spectrum, not to line-like features, thus we compute the photon flux only using the position-independent prompt emission in \Eq{alpha spectrum} with $\alpha = \gamma$. Contrary to the extragalactic flux, the flux in the CGH region is sensitive to the assumed DM density profile due to the pronounced differences between cored and cuspy profiles close to the GC. For this reason we use the \HESS~bound in the CGH region to set a profile-independent bound on $\Lambda_i^a / \bar{J}^{1/4}$ with
\beq
\bar{J} \equiv \left( \int_\text{CGH} \ud b \, \ud \ell \right)^{-1} \int_\text{CGH} \ud b \, \ud \ell \int_\text{l.o.s.} \frac{\ud s}{r_\odot} \frac{\rho_\text{halo}(r(s, \psi(b, \ell)))}{\rho_\odot}
\eeq
the angular-averaged $J$ factor in the sky region of interest. Notice that this bound is truly profile-independent as long as position-dependent processes such as IC can be neglected. For reference, the value of $\bar{J}$ for a cuspy profile like NFW~\cite{Navarro:1995iw} and a cored profile like Burkert~\cite{Burkert:1995yz} computed with the functions in Ref.~\cite{Cirelli:2010xx} is
\begin{align}
\label{Jfactorz}
\bar{J}_\text{NFW} \approx 17.38 \ ,
&&&
\bar{J}_\text{Burkert} \approx 4.47 \ .
\end{align}
Our $95 \%$ CL bounds on $\Lambda_2^a$ are summarized in Table~\ref{LambdaBounds}, together with the respective limits on the DM lifetime $\tau_\text{DM}$ derived by considering all relevant decay rates.

The \FERMI~limits from the continuum gamma-ray flux prove to be stronger than the \HESS~bounds from gamma-ray lines (as also found by Ref.~\cite{Aisati:2015ova}), up to a factor of $8.7$ for the limit on $\tau_\text{DM}$ almost independently on the DM profile in the galaxy. While \HESS's current sensitivity on the partial decay width into channels with prompt photons is at the level of $\Gamma_\gamma^{-1} \sim 10^{28}$ s~\cite{Aisati:2015ova}, our bounds on $\tau_\text{DM} \gtrsim 10^{27}$ s are less stringent due to the fact we include all relevant decay channels. In other words, we constrain the full DM decay width rather than the partial width into channels with prompt photons, by also taking into account the important contribution of other decay modes.

\begin{table}[t]
\begin{center}
\scriptsize{
\begin{tabular}{>{\rule[-1.8mm]{0mm}{5mm}} c | c | c | c | c | c}
& & \multicolumn{2}{c |}{$\overline{\X} L H H H^\dagger$} & \multicolumn{2}{c}{$\overline{\X} \sigma_{\mu \nu} L W^{\mu \nu} H$}
\\
\hline
&
$a$
&
Min $\Lambda_1^a$ [$\times 10^{16}$ GeV]
&
Min $\tau_\text{DM}$ [$\times 10^{27}$ s]
&
Min $\Lambda_2^a$ [$\times 10^{16}$ GeV]
&
Min $\tau_\text{DM}$ [$\times 10^{27}$ s]
\\
\hline
\multirow{3}{*}{\parbox[t]{3.0cm}{\FERMI~continuum, DM signal only ($3 \sigma$)}}
&
$e$
&
$1.05$ 
&
$1.95$
&
$5.64$
&
$2.15$
\\
&
$\mu$
&
$1.03$
&
$1.82$
&
$5.56$
&
$2.03$
\\
&
$\tau$
&
$1.17$
&
$2.97$
&
$6.16$
&
$3.06$
\\
\hline
\multirow{3}{*}{\parbox[t]{3.0cm}{\FERMI~continuum, DM signal + background ($3 \sigma$)}}
&
$e$
&
$1.59$
&
$10.1$
&
$8.19$
&
$9.55$
\\
&
$\mu$
&
$1.50$
&
$8.05$
&
$7.78$
&
$7.78$
\\
&
$\tau$
&
$1.52$
&
$8.55$
&
$7.88$
&
$8.18$
\\
\hline
\multirow{3}{*}{\parbox[t]{3.0cm}{\HESS~gamma-ray line, CGH region ($95\%$ CL)}}
&
$e$
&
$-$
&
$-$
&
$2.20 \times \bar{J}^{1/4}$
&
$0.05 \times \bar{J}$
\\
&
$\mu$
&
$-$
&
$-$
&
$2.20 \times \bar{J}^{1/4}$
&
$0.05 \times \bar{J}$
\\
&
$\tau$
&
$-$
&
$-$
&
$2.20 \times \bar{J}^{1/4}$
&
$0.05 \times \bar{J}$
\\
\hline
\multirow{3}{*}{\parbox[t]{3.0cm}{\HESS~gamma-ray line, extragalactic ($95\%$ CL)}}
&
$e$
&
$-$
&
$-$
&
$4.78$
&
$1.10$
\\
&
$\mu$
&
$-$
&
$-$
&
$4.78$
&
$1.10$
\\
&
$\tau$
&
$-$
&
$-$
&
$4.78$
&
$1.10$
\end{tabular}
}
\caption{\label{LambdaBounds}\em
Gamma-ray bounds on the new-physics scale $\Lambda_{1, 2}^a$ defined in \Eq{Lambdaia} and on the DM lifetime $\tau_\text{DM}$, separately for the two operators $(\Lambda_1^a)^{-2} \overline{\X} L^a H H H^\dagger$ and $(\Lambda_2^a)^{-2} \overline{\X} \sigma_{\mu \nu} L^a W^{\mu \nu} H$ and for each lepton flavor $a = e, \mu, \tau$. Both operators are constrained by the \FERMI~measurement of the isotropic diffuse gamma-ray flux, which is used here to derive a conservative bound considering the DM signal alone, and a realistic bound considering DM signal + background. The dipole operator induces a gamma-ray line-like feature in the photon spectrum, and thus is also constrained by \HESS~searches of gamma-ray lines in the CGH and extragalactic regions. The DM-profile dependence of the bounds from the CGH region is factored in the $\bar{J}$ factor, values for which are given in \Eq{Jfactorz} for two example density profiles. All other bounds are reasonably independent of the DM profile in the halo.}
\end{center}
\end{table}

\begin{figure}[t]
\centering
\includegraphics[width=0.325\textwidth]{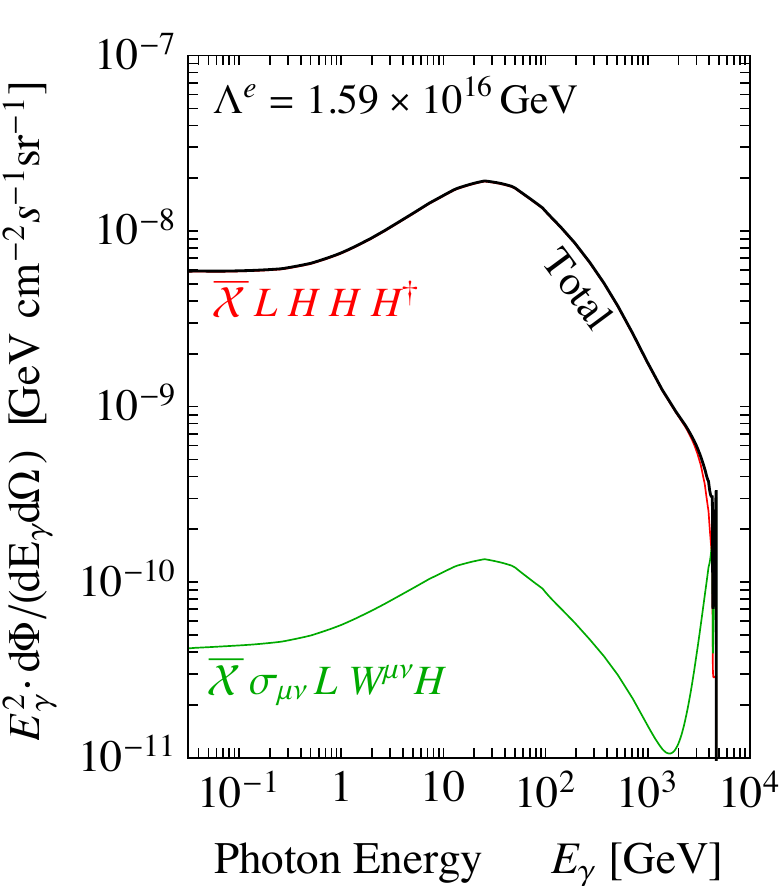}
\includegraphics[width=0.325\textwidth]{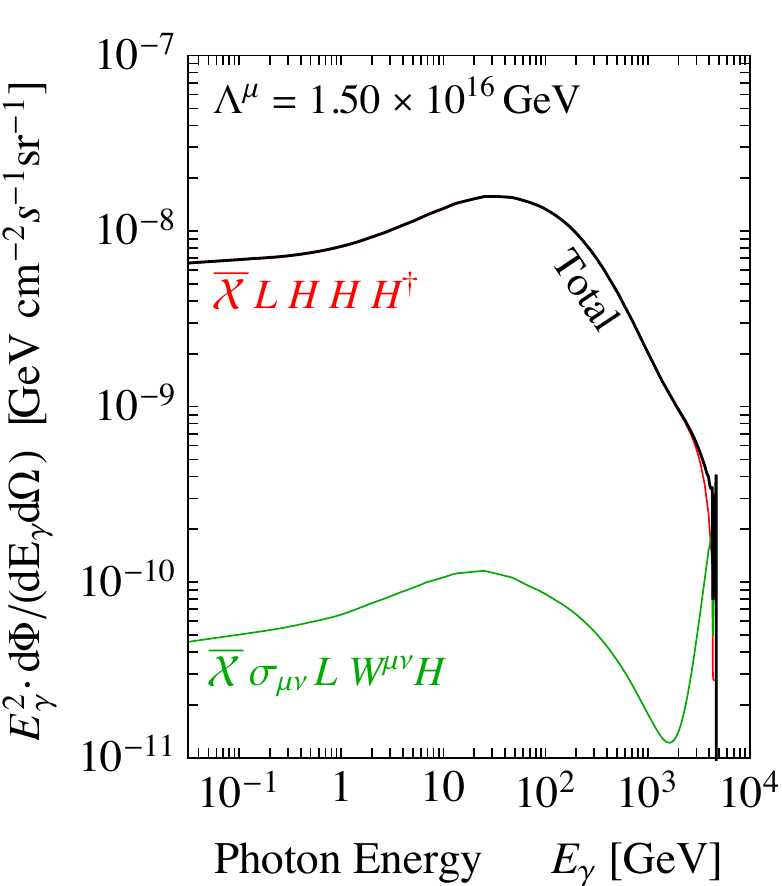}
\includegraphics[width=0.325\textwidth]{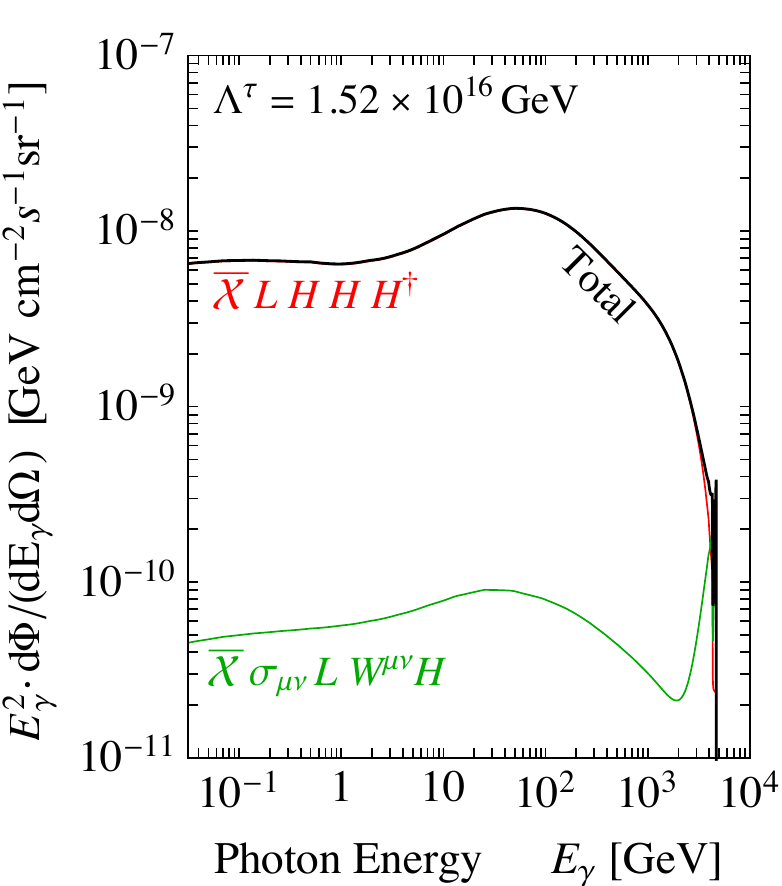}
\caption{\label{fig:line}\em
Isotropic gamma-ray flux due to DM decays induced by the operators $(\Lambda^a)^{-2} \overline{\X} L^a H H H^\dagger$ (red line) and $(\alpha_2 / 4 \pi) (\Lambda^a)^{-2} \overline{\X} \sigma_{\mu \nu} L^a W^{\mu \nu} H$ (green line), and their sum (black line). The suppression factor for the dipole coupling is expected from a radiative nature of the operator and causes the gamma-ray line-like signal to be dwarfed by the continuum photon flux. The three plots assume DM coupling to $a = e$ ({\bf left}), $a = \mu$ ({\bf center}), and $a = \tau$ ({\bf right}). The best-fitting value of $\Lambda^a$, adopted here to normalize the fluxes, is reported in the upper part of each plot.}
\end{figure}

The bounds shown in Table~\ref{LambdaBounds} were derived separately for the two operators $\overline{\X} L H H H^\dagger$ and $\overline{\X} \sigma_{\mu \nu} L W^{\mu \nu} H$, assuming only one was turned on at a time. However, in general, both operators are expected to arise in the effective theory description, and, if they are generated by the same physics at the scale $\Lambda$, we also expect their coefficients $c_1^a$ and $c_2^a$ to be somehow related. Since the dipole operator $\overline{\X} \sigma_{\mu \nu} L W^{\mu \nu} H$ is certainly generated at loop level, while $\overline{\X} L H H H^\dagger$ can conceivably originate at tree level, we can guess that $c_2^a \approx (\alpha_2 / 4 \pi) c_1^a$ with $\alpha_2 \approx 1/25$ as expected from the renormalization group evolution of the weak coupling if the new physics in the loop is at the GUT scale. Therefore, the prospects of detecting the gamma-ray line-like feature originating from the dipole operator are much worse than naively expected from the study of the operator alone. Fig.~\ref{fig:line} shows the gamma-ray flux due to DM decays induced by the operators $\frac{1}{\Lambda^2} \overline{\X} L H H H^\dagger$ (red line) and $\frac{\alpha_2}{4 \pi \Lambda^2} \overline{\X} \sigma_{\mu \nu} L W^{\mu \nu} H$ (green line), and their sum (black line). It is clear that the resulting line-like feature is much less visible against the continuum of photons than in our previous analysis considering just one operator. This result shows that analyses of gamma-ray line-like signatures of specific operators within an effective theory description should be accompanied by an assessment of the contribution to the continuum photon flux of all other operators that are expected in the effective theory.

\section{Conclusions}
\label{Conclusions}
Minimal Dark Matter (MDM)~\cite{Cirelli:2005uq} is a theoretical framework highly appreciated for its minimality and yet its predictivity. Contrary to many models where DM stability is imposed by hand through a global symmetry, MDM candidates are made stable on cosmological timescales by accidental symmetries occurring through a careful selection of the DM quantum numbers. When the cutoff of the model is taken to be the Planck scale, internal consistency conditions (the absence of Landau poles below the cutoff scale) and phenomenological constraints single out a fermionic $\SUd$ quintuplet and a scalar eptaplet as the only viable MDM candidates.

Recently, the MDM model was endangered by the discovery that the eptaplet decays quickly due to a previously overlooked dimension-$5$ operator~\cite{DiLuzio:2015oha}, and thus it is not a viable candidate; and by stringent gamma-ray line constraints in the Galactic Center, which do or almost do rule out the quintuplet, depending on the assumed DM density profile in the halo~\cite{Cirelli:2015bda, Garcia-Cely:2015dda}. In the light of these recent results, a critical reanalysis of MDM aiming at generalizing and extending this framework was in order.

This is the purpose of the present paper. After reviewing the MDM setup and its assumptions, we proposed two possible generalizations and studied their phenomenological implications. First, we found that MDM multiplets with a small enough hypercharge provide viable DM candidates, which possess small electric charges (the so-called millicharged DM) and are therefore absolutely stable. We discussed the case of millicharged singlet DM, and determined the thermal relic of triplets, quintuplets and eptaplets thus obtaining their mass. Interestingly, we found that a Dirac triplet is not constrained by the gamma-ray line searches that, for a cuspy DM halo profile, rule out a Wino (Majorana triplet) and the original MDM quintuplet.

Second, we proposed the possibility of lowering the Planck-scale cutoff for the original model of MDM quintuplet with zero hypercharge. As a consequence, the DM can decay by means of two dimension-$6$ operators which break the accidental symmetry and we can observe the signature of these decays in the gamma-ray sky. We found the cutoff to be constrained by \FERMI~data on the diffuse isotropic gamma-ray flux at about the GUT scale. We also discussed the constraints set by \HESS~on the gamma-ray line-like feature produced by the dipole operator, finding that the \FERMI~data set a stronger bound for a $10$ TeV DM. We also found that, when the dipole operator is assumed to be generated by loop processes, this line-like feature is completely dwarfed by the photon continuum induced by the other operator. Were a clear photon line from this candidate's annihilations to be soon detected, gamma-ray data could also be used to gain insight on the scale of new physics above the DM mass.

\section*{Acknowledgements}
We thank Giorgio Arcadi, Marco Cirelli, Guido D'Amico, Luca Di Luzio, Nicolao Fornengo, Ben Gripaios, Michele Redi, Filippo Sala, Luca Vecchi and Andrea Wulzer for fruitful discussions, and Michael Gustafsson and Gabrijela Zaharijas for useful communications. E.D.N.~was supported in part by DOE under Award Number DE-SC0009937 and by the MIUR-FIRB Grant RBFR12H1MW. M.N.~acknowledges support of the STFC grant ST/L000385/1, and King's College, Cambridge. P.P.~acknowledges support of the European Research Council project 267117 hosted by Universit\'e Pierre et Marie Curie-Paris 6, PI J.~Silk.

\appendix

\section{Trilinear couplings}
\label{app:trilinear}

With the method of Hilbert series it is possible to prove that, combining three instances of the same irreducible $SU(2)$ representation,
\begin{enumerate}
\item A unique invariant can be constructed for even isospin, and no invariant can be constructed for odd isospin.
\item A unique isospin triplet can be constructed for odd isospin, and no triplet can be constructed for even isospin.
\end{enumerate}
We only quote here the relevant steps, referring the reader to Ref.~\cite{Lehman:2015via} for further details.

The character function of an irreducible $SU(2)$ representation of isospin $I$ is given by
\begin{equation}
\chi_I (z) = \sum_{k = - I}^{+ I} z^{2k} = \frac{z^{2 I + 1} - z^{- 2 I - 1}}{z - z^{-1}} \ .
\end{equation}
From this expression we can construct the plethystic exponential 
\begin{equation}
P(X) = \exp \! \left( \sum_{r = 1}^{+ \infty} \frac{X^r \chi_I (z^r)}{r} \right) ,
\end{equation}
where $X$ is the object transforming under the desired irreducible representation with isospin $I$. In the following manipulations, $X$ should be thought of as a complex number with $|X| < 1$. We are interested in the trilinear couplings of $X$, thus we Taylor-expand $P(X)$ at third order in $X$:
\begin{equation}
P(X) = 1 + \chi_I (z) X + \frac{1}{2} \left( \chi^2_I (z) + \chi_I (z^2) \right) X^2 + \frac{1}{6} \left( \chi^3_I (z) + 3 \chi_I (z) \chi_I (z^2) + 2 \chi_I (z^3) \right) X^3 + \Op(X^4) \ .
\end{equation}
The number of invariants ($d_0$) and of triplets ($d_1$) is given respectively by
\begin{align}
d_0 &= \frac{1}{12 \pi i} \oint \frac{\ud z}{z} (1 - z^2) \left( \chi^3_I (z) + 3 \chi_I (z) \chi_I (z^2) + 2 \chi_I (z^3) \right) ,
\\
d_1 &= \frac{1}{12 \pi i} \oint \frac{\ud z}{z} (1 - z^2) \chi_1 (z) \left( \chi^3_I (z) + 3 \chi_I (z) \chi_I (z^2) + 2 \chi_I (z^3) \right) ,
\end{align}
where the line integrals are to be taken along a closed line arbitrarily close to the origin. The only relevant pole is at $z = 0$ for both integrals, and evaluating the residues yields
\begin{align}
d_0 &= \frac{1}{(6 I) !} \lim_{z \to 0} \frac{\ud^{1 + 6 I}}{\ud z^{1 + 6 I}} \left( \frac{1 - z^{12 (1 + I)} - z^{2 - 4 I} - z^{4 + 4 I} + z^{6 + 8 I} + z^{8 + 8 I} + z^{10 + 8 I}}{1 - z^4 - z^6 + z^{10}} \right) ,
\\
d_1 &= \frac{1}{(2+6 I) !} \lim_{z \to 0} \frac{\ud^{3 + 6 I}}{\ud z^{3 + 6 I}} \left( \frac{1 - z^{12 (1 + I)} - z^{2 - 4 I} - z^{4 + 4 I} + z^{6 + 8 I} + z^{8 + 8 I} + z^{10 + 8 I}}{1 - z^2 - z^4 + z^{6}} \right) .
\end{align}
We verified that this results in
\begin{align}
d_0 = 
\begin{cases}
1 \text{ for even $I$},
\\
0 \text{ for odd $I$},
\end{cases}
&&& 
d_1 = 
\begin{cases}
0 \text{ for even $I$},
\\
1 \text{ for odd $I$},
\end{cases}
\end{align}
for $I \leqslant 400$.

\section{MDM quintuplet decay rates}
\label{app:rates}

\subsection{$\overline{\X} L H H H^\dagger$}
As explained in Sec.~\ref{Decaying MDM}, this operator induces the DM decay modes listed in \Eq{decaymodes1}. We give here detailed analytic expressions for these decay rates. We compute explicitly the $\DM \to \nu h, \nu h h, \nu h h h$ decay rates and then derive all other rates by applying the Equivalence Theorem.

Our analytic computation of the $4$-body phase space (approximating all final states as massless) is described in Appendix \ref{app:phasespace}. When computing decay rates into final states with $n$ identical particles, we consider the $n!$ identical diagrams contributing to the scattering amplitude, and the $1/n!$ phase space reduction factor to prevent double-counting identical configurations.

DM decays into a neutrino plus Higgses are given by
\begin{multline}
\frac{c_1^a}{\Lambda^2} \overline{\X} L^a H H H^\dagger + \hc \supset - \frac{c_1^a}{4 \sqrt{3} \Lambda^2} \overline{\DM} \nu^a_\LH (v+h)^3 + \hc
\\
\supset
- \frac{3 c_1^a v^2}{4 \sqrt{3}\Lambda^2} \overline{\DM} \nu^a_\LH h
- \frac{3 c_1^a v}{4 \sqrt{3} \Lambda^2} \overline{\DM} \nu^a_\LH h h
- \frac{c_1^a}{4 \sqrt{3} \Lambda^2} \overline{\DM} \nu^a_\LH h h h + \hc
\end{multline}
with $v = 246$ GeV. The relevant polarization sum entering the spin-averaged squared matrix element is
\begin{equation}
\frac{1}{2} \sum_{s,r} \left| \bar{u}^s (p_\nu) P_\RH u^r(p_\X) \right|^2 = p_\nu \cdot p_\X = \mDM E_\nu \ ,
\end{equation}
where in the last equality we set ourselves in the rest frame of the decaying particle.

\subsubsection{$\DM \to \nu h $}

\begin{align}
\frac{\ud \Gamma \left(\DM \to \nu^a h \right)}{\ud E_\nu \, \ud E_h} & = \frac{3 \left| c_1^a \right|^2 v^4}{256 \pi \Lambda^4} E_\nu \, \delta (E_\nu -\tfrac{\mDM}{2}) \delta (E_h - \tfrac{\mDM}{2})
\\
\Gamma \left(\DM \to \nu^a h\right) & = \frac{3 \left| c_1^a \right|^2 v^4}{512 \pi \Lambda^4} \mDM
\end{align}

\subsubsection{$\DM \to \nu h h $}

\begin{align}
\frac{\ud \Gamma \left(\DM \to \nu^a h h \right)}{\ud E_\nu \, \ud E_h} & = \frac{3 \left| c_1^a \right|^2 v^2}{512 \pi^3 \Lambda^4} E_\nu \ , \qquad \frac{\mDM}{2} \leqslant E_\nu + E_h, \ E_\nu \leqslant \frac{\mDM}{2}, \ E_h \leqslant \frac{\mDM}{2}
\\
\frac{\ud \Gamma \left(\DM \to \nu^a h h \right)}{\ud E_\nu} & = \frac{3 \left| c_1^a \right|^2 v^2}{512 \pi^3 \Lambda^4} E_\nu^2 \ , \qquad E_\nu \leqslant \frac{\mDM}{2}
\\
\frac{\ud \Gamma \left(\DM \to \nu^a h h \right)}{\ud E_h} & = \frac{3 \left| c_1^a \right|^2 v^2}{1024 \pi^3 \Lambda^4} E_h (\mDM - E_h) \ , \qquad E_h \leqslant \frac{\mDM}{2}
\\
\Gamma \left(\DM \to \nu^a h h\right) & = \frac{\left| c_1^a \right|^2 v^2}{4096 \pi^3 \Lambda^4} \mDM^3
\end{align}

\subsubsection{$\DM \to \nu h h h$}

\begin{align}
\frac{\ud \Gamma \left(\DM \to \nu^a h h h \right)}{\ud E_\nu \, \ud E_h} & = \frac{\left| c_1^a \right|^2}{8192 \pi^5 \Lambda^4} \times
\begin{cases}
4 E_\nu^2 E_h & E_\nu + E_h \leqslant \frac{\mDM}{2}, \ E_\nu \leqslant \frac{\mDM}{2}, \ E_h \leqslant \frac{\mDM}{2}
\\
E_\nu (\mDM - 2 E_\nu) (\mDM - 2 E_h) & E_\nu + E_h \geqslant \frac{\mDM}{2}, \ E_\nu \leqslant \frac{\mDM}{2}, \ E_h \leqslant \frac{\mDM}{2}
\end{cases}
\\
\frac{\ud \Gamma \left(\DM \to \nu^a h h h \right)}{\ud E_\nu} & = \frac{\left| c_1^a \right|^2}{16384 \pi^5 \Lambda^4} \mDM E_\nu^2 (\mDM - 2 E_\nu) \ , \qquad E_\nu \leqslant \frac{\mDM}{2}
\\
\frac{\ud \Gamma \left(\DM \to \nu^a h h h \right)}{\ud E_h} & = \frac{\left| c_1^a \right|^2}{49152 \pi^5 \Lambda^4} \mDM E_h (\mDM - E_h) (\mDM - 2 E_h) \ , \qquad E_h \leqslant \frac{\mDM}{2}
\\
\Gamma \left(\DM \to \nu^a h h h\right) & = \frac{ \left| c_1^a \right|^2}{1572864 \pi^5 \Lambda^4} \mDM^5
\end{align}

\subsubsection{Other channels}
Decay rates into final states with a (left-handed) fermion $f = \ell, \nu$ and longitudinal gauge bosons $V = Z_\LH, W_\LH^\pm$ are computed using the Equivalence Theorem. The following proportionality relations are found with channels with one neutrino and $n$ Higgs bosons in the final state:
\beq\label{c1equivalence}
\frac{\ud \Gamma \left( \text{mode} \right)}{\ud E_f \, \ud E_V} = g_n \! \left( \text{mode} \right) \left. \frac{\ud \Gamma \left(\DM \to \nu h^n \right)}{\ud E_\nu \, \ud E_h} \right|_{\substack{E_\nu = E_f \\ E_h = E_V}} \ ,
\eeq
with
\begin{align}
g_1 \! \left( \DM \to \nu Z_\LH \right) &= 1/9 \ ,
&
g_1 \! \left( \DM \to \ell W_\LH \right) &= 8/9 \ ,
\\
g_2 \! \left( \DM \to \nu Z_\LH Z_\LH \right) &= 1/9 \ , \rule{0mm}{6mm}
&
g_2 \! \left( \DM \to \nu Z_\LH h \right) &= 2/9 \ ,
\\
g_2 \! \left( \DM \to \nu W_\LH^+ W_\LH^- \right) &= 8/9 \ ,
&
g_2 \! \left( \DM \to \ell W_\LH^+ h \right) &= 16/9 \ ,
\\
g_3 \! \left( \DM \to \nu Z_\LH Z_\LH Z_\LH \right) &= 1 \ , \rule{0mm}{6mm}
&
g_3 \! \left( \DM \to \nu Z_\LH W_\LH^+ W_\LH^- \right) &= 8/3 \ ,
\\
g_3 \! \left( \DM \to \nu Z_\LH Z_\LH h \right) &= 1/3 \ ,
&
g_3 \! \left( \DM \to \nu W_\LH^+ W_\LH^- h \right) &= 8/3 \ ,
\\
g_3 \! \left( \DM \to \nu Z_\LH h h \right) &= 1/3 \ ,
&
g_3 \! \left( \DM \to \ell Z_\LH Z_\LH W_\LH^+ \right) &= 8/3 \ ,
\\
g_3 \! \left( \DM \to \ell W_\LH^+ W_\LH^+ W_\LH^- \right) &= 8/3 \ ,
&
g_3 \! \left( \DM \to \ell W_\LH^+ h h \right) &= 8/3 \ .
\end{align}
These values take into account the appropriate $n!$ factors in the decay rate due to the presence of indistinguishable particles in the final state, both on the left and right-hand side of \Eq{c1equivalence}. As per assumptions of the Equivalence Theorem, these relations hold in the high-energy limit where the DM particle is much heavier than its decay products, which is true in this case given \Eq{quintupletmass}. Only longitudinal gauge bosons contribute significantly to the rate in this limit.

\subsection{$\overline{\X} \sigma_{\mu \nu} L W^{\mu \nu} H$}
From the second operator we get two terms inducing $\DM$ decay:
\begin{multline}
\frac{c_2^a}{\Lambda^2} \overline{\X} \sigma_{\mu \nu} L^a W^{\mu \nu} H + \hc \supset
- \frac{c_2^a}{2 \sqrt{6} \Lambda^2} \left(
2 \, \overline{\DM} \sigma^{\mu \nu} \ell^a_\LH W^3_{\mu \nu} \phi^+
+ \sqrt{2} \, \overline{\DM} \sigma^{\mu \nu} \nu^a_\LH W^-_{\mu \nu} \phi^+
\right. \\
\left.
- \overline{\DM} \sigma^{\mu \nu} \ell^a_\LH W^+_{\mu \nu} (h + v + i \phi^0)
+ \sqrt{2} \, \overline{\DM} \sigma^{\mu \nu} \nu^a_\LH W^3_{\mu \nu} (h + v + i \phi^0)
\right)
+ \hc
\end{multline}

The most relevant decays induced by this operator are $\DM \to \ell W, \nu Z, \nu \gamma, \ell W h, \nu Z h, \nu \gamma h$ (see Sec.~\ref{Decaying MDM}). The relevant polarization sum entering the spin-averaged squared matrix element of the processes $\DM \to f V (h)$ (with $f$ a fermion and $V$ a vector boson) is
\begin{equation}
\frac{1}{2} \sum_q \sum_{s,r} \left| \bar{u}^s (p_f) \sigma_{\mu\nu} P_\RH u^r(p_\X) p_V^\mu \varepsilon^{\nu *}_q(p_V) \right|^2 = 4 (p_\X \cdot p_V) (p_f \cdot p_V) \ .
\end{equation}

\subsubsection{$\DM \to \ell W$}
\begin{align}
\overline{\left| \Mel(\DM \to \ell^a W) \right|^2} & = \frac{v^2 \left| c^a_2 \right|^2}{3 \Lambda^4} \mDM^3 E_W
\\
\frac{\ud \Gamma \left( \DM \to \ell^a W \right)}{\ud E_\ell \, \ud E_W} & = \frac{v^2 \left| c^a_2 \right|^2}{48 \pi \Lambda^4} \mDM^2 E_W \, \delta (E_\ell - \tfrac{\mDM}{2}) \delta (E_W - \tfrac{\mDM}{2})
\\
\Gamma \left( \DM \to \ell^a W \right) & = \frac{v^2 \left| c^a_2 \right|^2}{96 \pi \Lambda^4} \mDM^3
\end{align}

\subsubsection{$\DM \to \nu Z$}
\begin{align}
\overline{\left| \Mel(\DM \to \nu^a Z) \right|^2} & = \frac{2 v^2 \left| c^a_2 \right|^2 \cos^2\theta_\text{W}}{3 \Lambda^4} \mDM^3 E_Z
\\
\frac{\ud \Gamma \left( \DM \to \nu^a Z \right)}{\ud E_\nu \, \ud E_Z} & = \frac{v^2 \left| c^a_2 \right|^2 \cos^2\theta_\text{W}}{24 \pi \Lambda^4} \mDM^2 E_Z \, \delta (E_\nu - \tfrac{\mDM}{2}) \delta (E_Z - \tfrac{\mDM}{2})
\\
\Gamma \left( \DM \to \nu^a Z \right) & = \frac{v^2 \left| c^a_2 \right|^2 \cos^2\theta_\text{W}}{48 \pi \Lambda^4} \mDM^3
\end{align}

\subsubsection{$\DM \to \nu \gamma$}
\begin{align}
\overline{\left| \Mel(\DM \to \nu^a \gamma) \right|^2} & = \frac{2 v^2 \left| c^a_2 \right|^2 \sin^2\theta_\text{W}}{3 \Lambda^4} \mDM^3 E_\gamma
\\
\frac{\ud \Gamma \left( \DM \to \nu^a \gamma \right)}{\ud E_\nu \, \ud E_Z} & = \frac{v^2 \left| c^a_2 \right|^2 \sin^2\theta_\text{W}}{24 \pi \Lambda^4} \mDM^2 E_\gamma \, \delta (E_\nu - \tfrac{\mDM}{2}) \delta (E_\gamma - \tfrac{\mDM}{2})
\\
\Gamma \left( \DM \to \nu^a \gamma \right) & = \frac{v^2 \left| c^a_2 \right|^2 \sin^2\theta_\text{W}}{48 \pi \Lambda^4} \mDM^3
\end{align}

\subsubsection{$\DM \to \ell W h$}
\begin{align}
\overline{\left| \Mel(\DM \to \ell^a W h) \right|^2} & = \frac{\left| c^a_2 \right|^2}{3 \Lambda^4} \mDM^2 E_W (\mDM - 2 E_h)
\\
\frac{\ud \Gamma \left( \DM \to \ell^a W h \right)}{\ud E_\ell \, \ud E_W \, \ud E_h} & = \frac{\left| c^a_2 \right|^2}{192 \pi^3 \Lambda^4} \mDM E_W (\mDM - 2 E_h) \, \delta(E_\ell + E_W + E_h - \mDM)
\\
\frac{\ud \Gamma \left( \DM \to \ell^a W h \right)}{\ud E_\ell} & = \frac{\left| c^a_2 \right|^2}{1152 \pi^3 \Lambda^4} \mDM E_\ell^2 (3 \mDM - 2 E_\ell)
\\
\frac{\ud \Gamma \left( \DM \to \ell^a W h \right)}{\ud E_W} & = \frac{\left| c^a_2 \right|^2}{192 \pi^3 \Lambda^4} \mDM E_W^3
\\
\frac{\ud \Gamma \left( \DM \to \ell^a W h \right)}{\ud E_h} & = \frac{\left| c^a_2 \right|^2}{384 \pi^3 \Lambda^4} \mDM E_h (\mDM - E_h)(\mDM - 2 E_h)
\\
\Gamma \left( \DM \to \ell^a W h \right) & = \frac{\left| c^a_2 \right|^2}{12288 \pi^3 \Lambda^4} \mDM^5
\end{align}

\subsubsection{$\DM \to \nu Z h$}
\begin{align}
\overline{\left| \Mel(\DM \to \nu^a Z h) \right|^2} & = \frac{2 \left| c^a_2 \right|^2 \cos^2\theta_\text{W}}{3 \Lambda^4} \mDM^2 E_Z (\mDM - 2 E_h)
\\
\frac{\ud \Gamma \left( \DM \to \nu^a Z h \right)}{\ud E_\nu \, \ud E_Z \, \ud E_h} & = \frac{\left| c^a_2 \right|^2 \cos^2\theta_\text{W}}{96 \Lambda^4} \mDM E_Z (\mDM - 2 E_h) \, \delta(E_\nu + E_Z + E_h - \mDM)
\\
\frac{\ud \Gamma \left( \DM \to \nu^a Z h \right)}{\ud E_\nu} & = \frac{\left| c^a_2 \right|^2 \cos^2\theta_\text{W}}{576 \Lambda^4} \mDM E_\nu^2 (3 \mDM - 2 E_\nu)
\\
\frac{\ud \Gamma \left( \DM \to \nu^a Z h \right)}{\ud E_W} & = \frac{\left| c^a_2 \right|^2 \cos^2\theta_\text{W}}{96 \Lambda^4} \mDM E_Z^3
\\
\frac{\ud \Gamma \left( \DM \to \nu^a Z h \right)}{\ud E_h} & = \frac{\left| c^a_2 \right|^2 \cos^2\theta_\text{W}}{192 \Lambda^4} \mDM E_h (\mDM - E_h)(\mDM - 2 E_h)
\\
\Gamma \left( \DM \to \nu^a Z h \right) & = \frac{\left| c^a_2 \right|^2 \cos^2\theta_\text{W}}{6144 \pi^3 \Lambda^4} \mDM^5
\end{align}

\subsubsection{$\DM \to \nu \gamma h$}
\begin{align}
\overline{\left| \Mel(\DM \to \nu^a \gamma h) \right|^2} & = \frac{2 \left| c^a_2 \right|^2 \sin^2\theta_\text{W}}{3 \Lambda^4} \mDM^2 E_\gamma (\mDM - 2 E_h)
\\
\frac{\ud \Gamma \left( \DM \to \nu^a \gamma h \right)}{\ud E_\nu \, \ud E_\gamma \, \ud E_h} & = \frac{\left| c^a_2 \right|^2 \sin^2\theta_\text{W}}{96 \Lambda^4} \mDM E_\gamma (\mDM - 2 E_h) \, \delta(E_\nu + E_\gamma + E_h - \mDM)
\\
\frac{\ud \Gamma \left( \DM \to \nu^a \gamma h \right)}{\ud E_\nu} & = \frac{\left| c^a_2 \right|^2 \sin^2\theta_\text{W}}{576 \Lambda^4} \mDM E_\nu^2 (3 \mDM - 2 E_\nu)
\\
\frac{\ud \Gamma \left( \DM \to \nu^a \gamma h \right)}{\ud E_\gamma} & = \frac{\left| c^a_2 \right|^2 \sin^2\theta_\text{W}}{96 \Lambda^4} \mDM E_\gamma^3
\\
\frac{\ud \Gamma \left( \DM \to \nu^a \gamma h \right)}{\ud E_h} & = \frac{\left| c^a_2 \right|^2 \sin^2\theta_\text{W}}{192 \Lambda^4} \mDM E_h (\mDM - E_h)(\mDM - 2 E_h)
\\
\Gamma \left( \DM \to \nu^a \gamma h \right) & = \frac{\left| c^a_2 \right|^2 \sin^2\theta_\text{W}}{6144 \pi^3 \Lambda^4} \mDM^5
\end{align}

\subsubsection{Other channels}
Decay rates into final states with gauge bosons $V = Z, W^\pm$ are computed using the Equivalence Theorem. To avoid the shortcomings of the Theorem we check the computation in the Equivalent gauge~\cite{Wulzer:2013mza}. The following proportionality relations are found with channels with only Higgs bosons in the final state:
\begin{align}
\label{c2equivalence}
\ud \Gamma \left( \DM \to \ell^a Z W^+_\LH \right) &= 2 \, \ud \Gamma \left( \DM \to \nu^a Z h \right) ,
\\
\ud \Gamma \left( \DM \to \ell^a \gamma W^+_\LH \right) &= 2 \, \ud \Gamma \left( \DM \to \nu^a \gamma h \right) ,
\\
\ud \Gamma \left( \DM \to \nu^a W^- W^+_\LH \right) &= 2 \, \ud \Gamma \left( \DM \to \ell^a W^+ h \right) ,
\\
\ud \Gamma \left( \DM \to \ell^a W^+ Z_\LH \right) &= \ud \Gamma \left( \DM \to \ell^a W^+ h \right) ,
\\
\ud \Gamma \left( \DM \to \nu^a Z Z_\LH \right) &= \ud \Gamma \left( \DM \to \nu^a Z h \right) ,
\\
\ud \Gamma \left( \DM \to \nu^a \gamma Z_\LH \right) &= \ud \Gamma \left( \DM \to \nu^a \gamma h \right) .
\end{align}
These values take into account the appropriate $n!$ factors in the decay rate due to the presence of indistinguishable particles in the final state. \Eq{c2equivalence} holds in the high-energy limit with the DM particle much heavier than its decay products, which is true for MDM. Only longitudinal gauge bosons contribute significantly to the rate in this limit.

\section{Four-body phase space for decaying MDM}
\label{app:phasespace}
Here we compute the phase space for a decay process into four massless particles in the assumption the scattering matrix element depends on up to two final momenta.

The phase space for four final particles with momenta $p_1$, $p_2$, $p_3$, and $p_4$ is
\beq
\ud \Phi^{(4)} = (2 \pi)^4 \delta^{(4)}(P^\mu - p_1^\mu - p_2^\mu - p_3^\mu - p_4^\mu) \frac{\ud^3 p_1}{(2 \pi)^3 2 E_1} \frac{\ud^3 p_2}{(2 \pi)^3 2 E_2} \frac{\ud^3 p_3}{(2 \pi)^3 2 E_3} \frac{\ud^3 p_4}{(2 \pi)^4 2 E_4} \ ,
\eeq
with $P^\mu$ the initial momentum, $E_i \equiv | \bol{p}_i |$ for $i = 1, 2, 3, 4$, and the customary relativistic state normalization $\langle \bol{p}' | \bol{p} \rangle = 2 E_i \, (2 \pi)^3 \delta^{(3)}(\bol{p} - \bol{p}')$ for momentum eigenstates of massless particles is adopted. We can now insert the identity in the form
\beq
\begin{split}
1 & = \int \frac{\ud s}{2 \pi} \, (2 \pi) \delta(s - q^2) \int \frac{\ud^4 q}{(2 \pi)^4} \, (2 \pi)^4 \delta^{(4)}(q^\mu - p_3^\mu - p_4^\mu) \theta(q^0)
\\
& = \int \frac{\ud s}{2 \pi} \int \frac{\ud^3 q}{(2 \pi)^3 2 E_{\bol{q}}} \, (2 \pi)^4 \delta^{(4)}(q^\mu - p_3^\mu - p_4^\mu) \ ,
\end{split}
\eeq
where in the second line we integrated over $q^0$ and defined $E_{\bol{q}} \equiv \sqrt{s + \bol{q}^2}$. Thus we have
\beq
\ud \Phi^{(4)} = \frac{\ud s}{2 \pi} \, \ud \Phi^{(3)}(P; p_1, p_2, q) \, \ud \Phi^{(2)}(q; p_3, p_4) \ ,
\eeq
where
\begin{align}
\ud \Phi^{(3)}(P; p_1, p_2, q) & = (2 \pi)^4 \delta^{(4)}(P^\mu - p_1^\mu - p_2^\mu - q^\mu) \frac{\ud^3 p_1}{(2 \pi)^3 2 E_1} \frac{\ud^3 p_2}{(2 \pi)^3 2 E_2} \frac{\ud^3 q}{(2 \pi)^3 2 E_{\bol{q}}} \ ,
\\
\ud \Phi^{(2)}(q; p_3, p_4) & = (2 \pi)^4 \delta^{(4)}(q^\mu - p_3^\mu - p_4^\mu) \frac{\ud^3 p_3}{(2 \pi)^3 2 E_3} \frac{\ud^3 p_4}{(2 \pi)^4 2 E_4} \ .
\end{align}
If the integrand does not depend on $p_3$ and $p_4$, the Lorentz-invariant two-body phase space can be integrated in the reference frame where $\bol{q} = \bol{0}$ thus yielding the well known result $\int \ud \Phi^{(2)}(q; p_3, p_4) = 1 / (8 \pi)$ for massless particles.

The remaining three-body phase space can be reduced in the following way. First we integrate away the three-momentum conservation delta function by performing the integral in $\ud^3 q$:
\beq
\ud \Phi^{(3)} = (2 \pi) \delta(E - E_1 - E_2 - E_{\bol{q}}) \frac{\bol{p}_1^2 \, \ud p_1 \, \ud\cos\theta_1 \, \ud\varphi_1}{(2 \pi)^3 2 E_1} \frac{\bol{p}_2^2 \, \ud p_2 \, \ud\cos\theta_2 \, \ud\varphi_2}{(2 \pi)^3 2 E_2} \frac{1}{2 E_{\bol{q}}} \ ,
\eeq
where $E$ is the DM energy. For decay of a scalar or unpolarized (spin-averaged) state, the distribution of the final state particles is isotropic and therefore we may integrate over two angles parametrizing rotations of the system as a whole, say $\theta_1$ and $\phi_1$. We define the polar angle $\theta_2$ relative to the direction of $\bol{p}_1$, so that the azimuthal angle $\varphi_2$ describes overall rotations of the system about $\bol{p}_1$ and therefore it can be trivially integrated over. We thus have
\beq
\ud \Phi^{(3)} = \frac{1}{4 (2 \pi)^3} \delta(E - E_1 - E_2 - E_{\bol{q}}) \, \frac{\bol{p}_1^2 \, \ud p_1}{E_1} \, \frac{\bol{p}_2^2 \, \ud p_2 \, \ud\cos\theta_2}{E_2} \, \frac{1}{E_{\bol{q}}} \ .
\eeq
Remembering now that $\delta(f(x)) = \sum_i \delta(x - x_i) / | f'(x_i) |$ where $x_i$ are the zeroes of $f(x)$, we can integrate away the delta function by performing the integral on $\cos\theta_2$ in the DM rest frame, where $\bol{q} = \bol{p}_1 + \bol{p}_2$ and $E = \mDM$:
\begin{multline}
\int_{-1}^{+1} \ud\cos\theta_2 \, \delta \Big( \mDM - E_1 - E_2 - \sqrt{s + E_1^2 + E_2^2 + E_1 E_2 \cos\theta_2} \Big)
\\
= \frac{E_{\bol{q}}}{E_1 E_2} \int_{-1}^{+1} \ud\cos\theta_2 \, \delta \Big( \cos\theta_2 - \frac{(\mDM - E_1 - E_2)^2 - (s + E_1^2 + E_2^2)}{2 E_1 E_2} \Big)
\\
= \frac{E_{\bol{q}}}{E_1 E_2} \, \theta \big( (\mDM - 2 E_1) (\mDM - 2 E_2) - s \big) \, \theta \big( s - \mDM (\mDM - 2 (E_1 + E_2)) \big) \ .
\end{multline}
The two theta functions appearing in the last line are the result of integrating the delta function and ensure that its argument lies within the integration support or otherwise the integral vanishes. Finally we can perform the integral over $s$,
\begin{multline}
\ud \Phi^{(4)} = \frac{\ud E_1 \, \ud E_2}{512 \pi^5} \int_0^\infty \ud s \, \theta \big( (\mDM - 2 E_1) (\mDM - 2 E_2) - s \big) \, \theta \big( s - \mDM (\mDM - 2 (E_1 + E_2)) \big)
\\
= \frac{\ud E_1 \, \ud E_2}{512 \pi^5} \times
\begin{cases}
(\mDM - 2 E_1) (\mDM - 2 E_2) & E_1 + E_2 \geqslant \mDM/2
\\
4 E_1 E_2 & E_1 + E_2 \leqslant \mDM/2
\end{cases}
\end{multline}
where $E_1$ and $E_2$ are bound to be smaller than $M/2$.

\end{document}